# ϒ (9.46 GeV) and the gluon discovery
# (a critical recollection of PLUTO results)*

Bruno R. Stella[1,a] and Hans-Jürgen Meyer[2,b]

[1] Department of Physics of Roma Tre University and INFN; Rome, Italy

[2] formerly at Department of Physics, Siegen University; Siegen, Germany

**Abstract.** The hadronic decays of ϒ(9.46 GeV) were first studied by the PLUTO experiment at the DORIS $e^+e^-$ storage ring (DESY). With the aim of determining the contribution of PLUTO to the discovery of the gluon, as members of this former collaboration we have reconsidered all the scientific material produced by PLUTO in 1978 and the first half of 1979. It results clearly that the experiment demonstrated the main decay of the ϒ(9.46 GeV) resonance to be mediated by 3 gluons, by providing evidence for the agreement of this hypothesis with average values and differential distributions of all possible experimental variables and by excluding all other possible alternative models.

Jettiness resulted evident by the average transverse momentum $<p_T>$ with respect to the event thrust axis, which was the same as experimentally observed by PLUTO itself at nearby continuum c.m.s. energies for 2-quark jet events. On the contrary, the average sphericity $<S>$ and more topological variables as well as the momentum distribution showed a net difference with the same data comparison, a result compatible with jettiness only in case of more than 2 jets. Flatness as consequence of a 3-body decay (therefore 3 jets) was indicated by the low average momentum out of the event plane $<p_{out}>$, altogether a result being independent of models. The charged multiplicity was observed to be larger than in the continuum and in case of simulated 3 gluon jets fragmenting like quarks, in the direction expected for gluon jets.

Moreover PLUTO measured in June 1979 the matrix element of the 3-gluon decay to be quantitatively as expected by QCD (even after hadronization, which does not obscure the perturbative predictions) and, having checked the possibility to correctly trace the fastest gluons direction, demonstrated the spin 1 nature of the gluon by excluding spin 0 and spin ½. The hadronization of the gluon like a quark jet, hypothesized in the 3-gluon jet Monte Carlo simulation, was perfectly compatible with the topological data at this energy and was shown to be an approximation at ≈10% level for the multiplicity; the true expected gluon fragmentation was needed to describe the inclusive distributions; this was the first experimental study of (identified) gluon jets.

In the following measurements at the PETRA storage ring, these results were confirmed by PLUTO and by three contemporaneous experiments by measuring at higher energies the gluon radiation ("bremsstrahlung"), the soft gluons by jet broadening, and the hard gluons by the emission of (now clearly visible) gluon jets by quarks. The gluon's spin 1 particle nature was also confirmed at PETRA. The PLUTO results on ϒ decays were confirmed both by contemporaneous experiments at DORIS (partially) and later (also partially) by more sophisticated detectors.

# 1  Introduction

Quantum Chromodynamics (QCD, the theory of strong interactions) and the gluon, the messenger of the strong ("color") force, were proposed (after early papers by Gell-Mann in 1962 [1] and 1964 [2] in which the gluon is mentioned for the first time as a neutral vector meson field) in the years 1970-1980 [3,4] (see also reviews [5,6]), in parallel and after the quark parton model was stabilized. A laboratory for studying QCD and gluons was proposed to be the next heavy narrow hadronic resonance [7-15]. In 1977 the ϒ(9.5 GeV) resonance was discovered at Fermilab [16,17] and its very narrow width (≈50 KeV) was found at DORIS (3-10 GeV $e^+e^-$ storage ring at DESY) by the experiments PLUTO [18] and DASP2 [19,20] in May 1978 and later by DHHM [21]. The first evidence for the abundant decay of ϒ into 3 gluons was reported by the PLUTO Collaboration, at Seminar, Schools and Conferences in Summer 1978 [23-31], as well as in publications [32-34]. These results were already quoted to be a strong hint toward the existence of the gluon ([27-29], the same repeated later in [44-46]). Further presentations followed at Winter [35,36] and Spring [37-40] Schools and meetings, and the cross sections were given in the thesis [41]. In June 1979 at the Geneva International Conference [42] the evidence for the ϒ decay into 3 gluons (with the partonic matrix element) was presented by PLUTO [43], mentioned also in [44,45], and the first evidence at PETRA (the new 10-48 GeV $e^+e^-$ storage ring at DESY) for quark jet broadening by gluon radiation was shown by TASSO [46] and also with more results by PLUTO [47]. At the following Lepton-Photon Symposium at FermiLab [48] PLUTO showed the step in R due to the production of the new quark b and confirmed the jet broadening [49] and the 3-gluon interpretation of the ϒ decay [50]. At this conference the evidence for three jet events (interpreted as gluon radiation by a q̄q pair) was shown by the TASSO, PLUTO, MARK-J and JADE experiments at PETRA [51-57] again confirming the existence of gluon jets now at a factor three larger energies. A review of the latest results from DESY, summarizing the evidence for gluons, both from ϒ→3-gluon decay and gluon bremsstrahlung, was presented by H. Schopper, director of the laboratory, at the Goa International Symposium in September 1979 [58].

As members of the PLUTO Collaboration, after more than thirty years we think it timely and worthwhile to recollect and recall in this article, and for a wider public, what PLUTO did in relation to the gluon discovery in the years 1978 and first half of 1979 and the confirmations obtained both at DORIS and at PETRA.[1]

In Chapter 2 we briefly summarize the related physics highlights preceeding the PLUTO experiment at DORIS; in Chapter 3 we sketch the PLUTO detector and the properties of the DORIS and PETRA storage rings, with a brief history of the machines and the detector; in Chapter 4 we outline the model simulations of the physical processes. In the main Chapter 5 we recollect the elements for the discovery of the ϒ→3-gluon decay: the ϒ resonance; inclusive dynamics; geometry (topology); exclusion of alternative models; exclusive 3-gluon dynamics and gluon hadronization (the first study of gluon jets). All with the aim to single out the sufficient and the necessary conditions to demonstrate the validity of the 3-gluon hypothesis (QCD). In Chapter 6 we cover the confirmations found at DORIS, especially by a more sophisticated detector (ARGUS), as well by CLEO at the CESR storage ring, Cornell, USA, the jet broadening found at PETRA and the most important confirmation for the gluon: the discovery of gluon bremsstrahlung. Finally we give a summary and draw the conclusions in Chapter 7.

---

[1] While writing this article a recent related historical review of P. Söding in this journal  [138] came to our knowledge. As that review mainly concentrated on the contribution to the gluon discovery by a different process, at a different collider (PETRA, at substantially higher c.m.s. energies), at a different (later)  time and with a different experiment, we consider the two articles to be complementary. Our view on the value of the ϒ decays in the search for evidence for gluons is different, as it is motivated also in Fig. 11 and footnote 12.

## 2  Prologue: The related physics in the years 1974-1978

The pointlike fractionally charged constituents of the elementary particles (partons or quarks) were hypothesized and found in the years 1964-1974. The last step, the number of quarks, was demonstrated experimentally by measuring R, the ratio $\sigma(e^+e^- \rightarrow hadrons) / \sigma(e^+e^- \rightarrow \mu^+\mu^-)$, a measurement of the sum of the square of the quark charges divided by the square of the muon's charge. A new quantum number ("color") was proposed [59] to justify the abundant production of hadrons and the high R value measured at the $e^+e^-$ colliders ADONE, CEA and SPEAR [60].

In 1974 a very narrow resonance was discovered [61-63], the J/$\psi$(3.1 GeV), recognised to be the ground state of a new q$\bar{q}$ resonance, of the "charm" quark, "charmonium" (c$\bar{c}$). In 1975 excited c$\bar{c}$ states were found, starting the field of charmonium spectroscopy (see episode 93 in [6]). Using the measured J/$\psi$ cross section and the J/$\psi \rightarrow e^+e^-$, $\mu^+\mu^-$ branching ratios, in non-relativistic potential models for quark binding, the charm quark was shown to have a charge ⅔ of the proton charge. The presence and charge of the new quark was also seen in $e^+e^-$ annihilations as a step in R outside the resonance region.

In 1975 Appelquist and Politzer [7,8] proposed (in analogy with the orthopositronium decay into 3 photons calculated with QED by Ore and Powell [64]) that a narrow q$\bar{q}$ resonance found in $e^+e^-$ annihilation (with the quantum numbers of the photon, as orthopositronium decaying into 3 γ's) should decay into 3 gluons, the supposed exchange particle of the strong interactions (with the same quantum numbers of the photon, plus "color": QCD is the name of the resulting theory). In the same year at SPEAR at 6.2 and 7.4 Gev c.m.s. energies the first "jets" of particles were identified in $e^+e^- \rightarrow$ q$\bar{q}$ annihilations [65] (later even extended to already show up as starting at 4.8 GeV c.m.s. energy [66,67]), a mechanism proposed [68-79] for the hadronization of quarks and gluons. (This means that a jet of ≈ 3 GeV or more is recognizable as a separate cluster of particles of limited transverse momentum of ≈ 0.36 GeV with respect to its mean longitudinal momentum: there is a threshold in energy to single out jets).

A new heavy lepton, the tau (τ) of mass ≈ 1.5 GeV, was also found in 1975 [80] as a third charged particle (after the electron and the muon). The τ, being heavy enough, can decay also into hadrons. Its existence implied the existence of a new heavy quark doublet ("beauty" and "truth" or "bottom" and "top") paired with it and the tau neutrino (according to the hypotheses of Glashow, Iliopulos and Maiani [81] and of Kobajashi and Maskawa [82]). The properties of the possible Q$\overline{Q}$ ground state were predicted in detail by Eichten and Gottfried in 1977 [83].

In 1976, Ellis, Gaillard and Ross [72] proposed that high energy quarks should radiate gluons (a 1$^-$ neutral massless colored particle) very much as in QED the electrons radiate photons. The q$\bar{q}$ pairs produced in $e^+e^-$ annihilations could radiate gluons (gluon bremsstrahlung) and those gluons could manifest themselves as a cascade of quarks and gluons and finally ordinary hadrons: jets again [72-77].

In 1976 the charmed mesons D and D* were discovered as bound states of a "light (u,d,s)" quark and the new charm quark (e.g. see episode 93 in [6]). In 1977 a ϒ(9.5 GeV) heavy resonance was discovered in an experiment at Fermilab of a proton beam striking a nuclear target [16,17], relatively narrow (±200 MeV, compatible with the resolution of the experiment) and seen in the $\mu^+\mu^-$ decay. Koller and Walsh [9-11] and in 1978-79 together with Krasemann, Zerwas and Krammer [13-15] and Fritzsch and Streng [12], proposed a test of QCD by looking for gluon jets in the decay of a heavy quark-antiquark bound state produced in $e^+e^-$ annihilations and calculated the gluons' or jets' (the forward product of their hadronization) angular distributions, estimating also multiplicities and momentum distributions of hadrons in a $Q\overline{Q} \to 3$ gluons $\to 3$ jets final state. When, at DESY, the Υ(9.46) was confirmed by PLUTO to be an extremely narrow state [18,84] but with abundant hadronic decays, it was clear that it was not decaying as a 'normal' hadronic resonance: it was a possible candidate for the proposed 3-gluon decay, manifesting themselves as three jets of hadrons, as every parton is supposed to do.

In order to support the 3-jet hypothesis, new topological quantities had to be defined for more inclusive information to evidence this final state, ideally calculable in QCD (i.e. safe from divergences at low energies and small angles, "infrared safe"). The measures had to work on events with two or more jets, even if broad and overlapping, for evidencing the final state. Many proposals were published in 1978-79: thrust [85,86], sphericity [65,71], spherocity [87], acoplanarity [76], triplicity [88], three-jettiness [89] and more.

## 3  The PLUTO Detector at DORIS and PETRA

The PLUTO detector [90,91] was originally designed (around 1970)  for experiments at the DORIS $e^+e^-$ storage ring for the energy range up  to 3.5 GeV  (see Fig. 1).  DORIS (Double Orbit Intersecting Storage Ring) was a colliding ring accelerator (initially with 480 bunches, $2\circ10^{30}$ cm$^{-2}$ sec$^{-1}$ peak luminosity), completed in 1973 and upgraded first to 9 GeV in 1977 and then, at the beginning of 1978, to 10.2 GeV for the Υ physics. The concept of PLUTO was a cylindrically arranged $4\pi$ detector (coaxial with the colliding beams) with almost 100% coverage for particles emerging from the interaction point. It was the first detector with a superconducting solenoid coil producing a homogeneous magnetic field of 1.69 Tesla for the inner track detector.

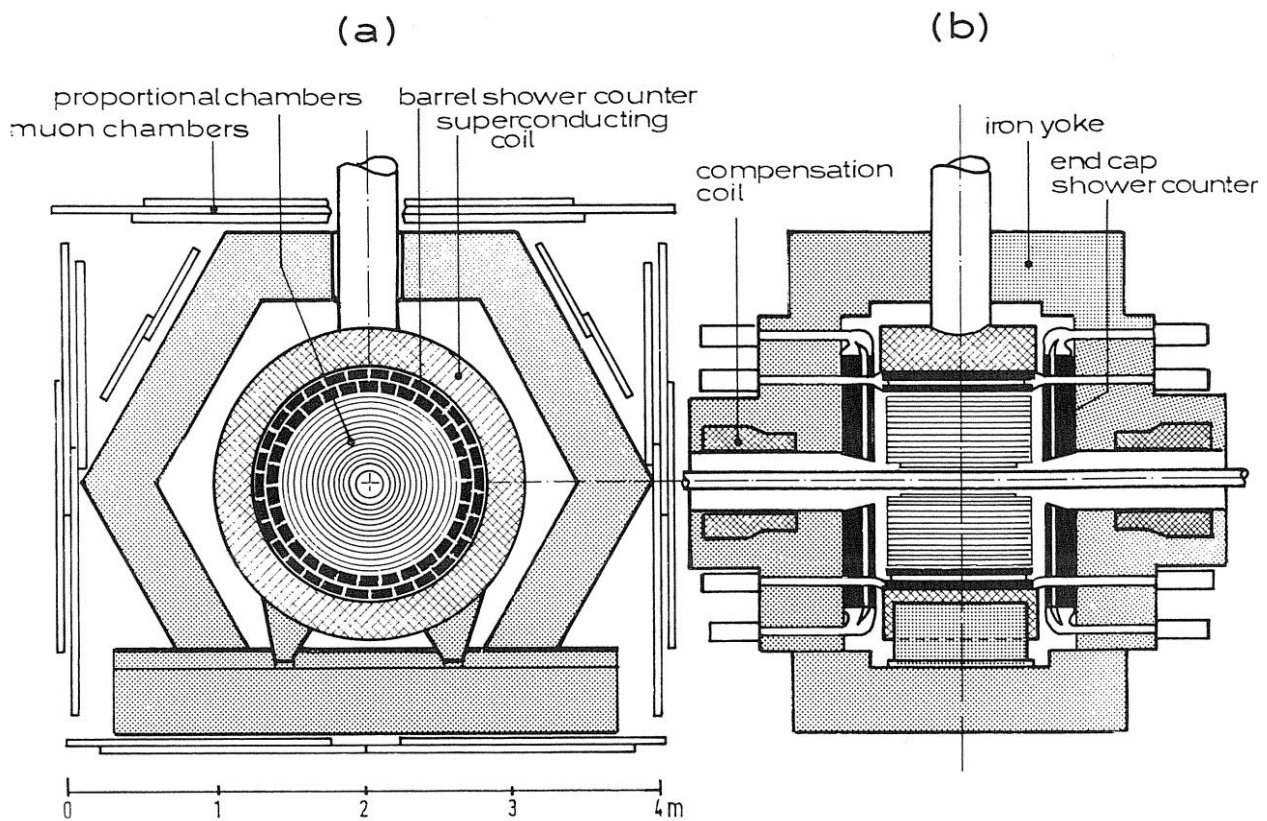

**Fig. 1.** Experimental set up of the PLUTO detector. Section a) perpendicular to and section b) containing the horizontal beam axis.

This first version of the detector consisted of a thin inner track gas detector with 10 layers of cylindrical proportional wire chambers (coverage 87% of 4π) for all charged particles and two layers of planar proportional chambers outside the flux return yoke (coverage 65% of 4π) for muon identification. Although planned for experiments in the "continuum energy range" of 3 - 5 GeV PLUTO was suitable for successful data takings and analyses for J/ψ, $\tau^+\tau^-$ and charmonium physics in the years 1974-1976.

DORIS was upgraded to 9 GeV in 1977  (now often called DORIS I), by conversion from a double storage ring with many bunches to a single storage ring with 2 bunches. In parallel, the PLUTO detector was upgraded by the addition of electromagnetic calorimetry, the so called barrel shower counter (8.6 radiation lengths) inside the coil and the endcap shower counters (10.5 radiation lengths) on both ends of the cylindrical volume. This gave  94% of 4π coverage for the detection of neutral particles, photon and electron identification and 92% for charged hadrons. Both shower counters were interspaced by wire chambers as to improve the spatial detection of neutral particles (see schematic view of the detector in Fig. 1). The momentum resolution for charged tracks was σ/p = 3% $p_T$ (p, $p_T$ in GeV, for p>3 GeV) and the shower-counter energy resolution (for E>1 GeV) was σ/E = 19%/√E (E in GeV) for the barrel and σ/E = 35%/√E (E in GeV) for the endcap counters. The detector was triggered either by the presence of tracks in the wire chambers, by sufficient detected energy in the calorimeter, or by a combination of the two.

Although planned for measurements in the continuum energy range of 7 - 9 GeV, the discovery of the new Υ(9.5±0.2) resonance in 1977 and the request by PLUTO to shift  the maximum DORIS I energy up to 10 GeV, the detector was ready again to perform unexpected data taking and analysis for resonance and jet physics in 1978. In April-May of this year the resonance was found in a narrow bin (9.46±0.01) by PLUTO [18] and DASP2 [19,20], (the latter only measuring the direction of particles). PLUTO accumulated an integrated  luminosity of 190 nb[-1] on the Υ resonance (9.45 < $E_{cm}$ < 9.47 GeV, corresponding to 1940 hadronic events) and 177 nb[-1] in the continuum (off: 9.3 GeV < $E_{cm}$ < 9.44 GeV, corresponding to 504 hadronic events [41]).

More radiofrequency cavities added to DORIS to increase the beam energy allowed DASP2 [92,93] and DHHM [21,22] detectors to find the Υ' at 10.02±0.01 GeV (an excitation already indicated in the Fermilab discovery). Then DORIS I reached its energy limit.

The November 1974 revolution (the discovery of the J/ψ and then of the new quark "charm" [61-63]) had persuaded DESY to convert PETRA (formerly Proton Electron Tandem Ring Accelerator) to a high energy $e^+e^-$ storage ring (up to 48 GeV c.m.s. energy), with high luminosity (2 bunches, 1.5∘$10^{31}$ cm[-2] sec[-1] peak luminosity), ready in November 1978.

The construction of PETRA lead to a third upgrade of PLUTO with the addition of muon chambers and forward spectrometers for the γγ physics, to be ready for data taking in the second half of 1978. The PLUTO detector was moved from DORIS I to PETRA and continued its good performance in data taking in the new PETRA energy range of 10 - 48 GeV. [A further upgrade of DORIS, DORIS II, was planned and realized meanwhile to continue and improve the exploration of the family of Υ ground state (bottomium) and its excitations, by the upgraded experiments DASP2 and DHHM (now LENA)  and the new detector ARGUS].

At the end the PLUTO detector had taken data at three different $e^+e^-$ storage rings spanning an energy range from about 3 to 32 GeV within a period of about 6 years (1974 - 1979), with three different machine conditions (different luminosity, optics, timing, background).  This made it a unique and successful experiment especially for aspects of QCD physics, particularly the analysis of quark and gluon jets with running $\alpha_S$, with the same detector being run by the same collaborating institutions (Aachen, DESY, Hamburg, Siegen, Wuppertal and later Bergen, Glasgow, Maryland, Tel Aviv) with membership increasing from about 30 to 80 members.  PLUTO was a prototype detector and collaboration for experiments at $e^+e^-$ storage rings.

The very good superconducting magnetic field and the availability of a good electromagnetic calorimeter were premium features compared with contemporary detectors without shower counters (for neutrals) or with  less precise charged particle information. PLUTO achieved the first DESY paper on jets in 1978 [32] , the discovery of the ϒ to be very narrow [18] and the  ϒ→ 3-gluon decay in 1978 [24,27,29,31-33], the confirmation of quark jet broadening (gluon radiation) [47,50] and together with TASSO, MARK-J and JADE the discovery of the gluon bremsstrahlung in 1979 [51-58][2].

## 4   The Monte Carlo Simulation of the physical processes

An important tool for modern particle physics experiments is the complete computer simulation of the experiment starting with the physical process, and including the detector, the electronic trigger and the selection of the events. In this way the response of the detector can be studied and optimized even before data taking starts, by making alternative hypotheses (more than one) about the physical process to be studied (of course, partly unknown) and later comparing them with data to decide which model performs best. Moreover the smearing of the physical variables due to the detector and to analysis methods can be appropriately corrected. Random variables are generated to simulate the randomness of the statistical fluctuations (hence the name "Monte Carlo"or shortly MC, first used by Enrico Fermi).

Since quarks and gluons ("partons") are not observable as free particles (due to their intense interactions, which confine them within a space roughly that of a proton ) and they "hadronize" (become normal hadrons) by fragmenting into more partons or materializing partons from the vacuum, which glue together by satisfying conservation of energy, momentum and other quantum numbers. Meanwhile the coupling "constant" $\alpha_{strong}$  (the intensity of the interaction) is actually not constant, but depends on  the scale, for example the momentum transfer during the interaction: a complicated process for both calculation and computer simulation.

The process of hadronization (the formation of jets of hadrons) is not calculable perturbatively in QCD (which means: it is non calculable by a converging series of decreasing terms), because at low momentum transfers the terms are increasing instead. This problem is addressed by phenomenological models. In 1978, an influential model for the quark "fragmentation" into jets of particles was proposed by Feynman and Field [99,100] called "independent fragmentation": each jet is an independent q → q' + meson cascade, repeating until there is insufficient energy to continue. Transverse momentum to the parton direction is limited and almost energy independent, longitudinal momenta are scaled with energy, strange-quark production from the quark-antiquark sea is suppressed and quantum numbers are locally compensated. This model, implemented and extended to  $e^+e^- \rightarrow q\bar{q}g$  by Hoyer et al. [101] and improved by Ali et al. [102] to include higher order QCD calculations and heavy quarks, was used successfully for years by experimentalists to simulate quark jets in Monte Carlo generators at  DORIS and PETRA energies. Gluon jets were expected to be different from quark jets, due to their different QCD "color" charge (8 color combinations for gluons, 3 for quarks) and from the "non abelian" nature of the strong interactions (self-coupling of gluons, not possible in QED for photons). The situation in August 1978 was summarized by de Rujula in his invited talk on "Jets" at the Tokyo Conference [77]. At that point the uncertainty about the nature of gluon jets was not small. Gluon jets were

---

supposed to be asymptotically different from quark jets, due to their self-coupling and their "flavor blindness"[3].

PLUTO simulated the 3-gluon decay by first accounting for the expected QCD dynamics (matrix element) and then fragmenting gluons (according to the just published Feynman and Field model [100]) mostly into $q\bar{q}$ pairs (as photons fragment into $e^+e^-$ pairs); quarks could also radiate gluons (as electrons and quarks radiate photons), initiating a cascade similar to the quark hadronization ending with hadronization at confinement energies (the low energy at which, if they split, their fragments bind together immediately due to the increased strength of the force).[4] In practice, the 3-gluon MC was adjusted as to describe the PLUTO experimental quark jets with the same momentum.

A full description of the PLUTO 3-gluon MC is given in [97,98]. The jets of hadrons in case of 2 quark jets at 9.4 GeV (continuum) were constructed according to the Field and Feynman independent fragmentation model [100] (see paragraph 4.2 of [98]). The detector, the trigger and the event selection were fully simulated.

The PLUTO 3-gluon MC approximation (gluon hadronization via a quark-antiquark pair of the same energy) was considered by the members of PLUTO itself to be very rough. However, it was confirmed experimentally by the Collaboration to be adequate at DORIS for the topological variables and the parton dynamics ($\Upsilon \to$3-gluon: see in the following Tab. 3 and Chap. 5.2) but not in single particle details (multiplicities, inclusive momentum distribution [36,41], strange particles [131] and baryon production [108], all different in the data at the level of ≈10% in the direction expected for QCD gluons: see Chap. 5.5). The 3-gluon MC model was also confirmed by PLUTO as a first approximation at PETRA at least up to 14 GeV c.m.s. including fragmentation (see Chap. 5.5). The fragmentation was later measured by JADE [108] in 1983 at c.m.s. energies in the range 22-36.4 GeV for bremsstrahlung gluons.

Also the 2 quark jets (fragmenting according to [100-102]) was a suitable model in principle for the $\Upsilon$ decay as for the continuum $e^+e^-$ annihilation.

The independent fragmentation model of Feynmann and Field [100-102], being the first quark jet model, had to be extended in 1981: more energy than expected by it was measured between jets [109]. This was improved by adding higher order terms, conservations, interferences and by extensions using the "cluster" model, and was accompanied by the "Lund" MC model [104-107] ("string" fragmentation: a string of the color force stretches between the generated partons and the fragmentation is no longer "independent" between final jets of hadrons). For a treatment of the hadronization models see Chap. 10 and 11 in [103].

As another alternative to the 3-gluon decay model one can also assume [41,98] the hadronic final state to be given simply by phase space, i.e. a constant matrix element. While phase space is not a realistic (dynamic) model, it stands for reactions which impose no particle correlation except for momentum and energy conservation. Besides the pion-only phase-space, two more phase-space generators have been used. One (P-PS) generates only pseudoscalar mesons (pions and kaons in the ratio 3:1). The other (P/V-PS) generates pseudoscalar and vector mesons in equal proportion (the most reasonable, according to T. Sjöstrand [Lund, private communication at that time]). The mean multiplicity of the particles produced in these phase-space Monte Carlo's was chosen to reproduce the observed multiplicities.

---

[3] In every gluon fragmentation into a pair of $q\bar{q}$ quarks the type of quark (named "flavor" quantum number) is compensated by its anti-quark, so that the different pairs of quarks are produced with almost the same probability (but for the different masses). This was something new, allowing particles made of quarks not existing in stable matter (e.g. "strange quarks") to appear in roughly 10% of the cases (we say then that strangeness is "suppressed" in nature). Nevertheless this modification of the zoo of particles in the final state of gluon jets was supposed to be small, due to the mass of the strange quark (and of the much heavier charm quark).

[4] The effects of radiative corrections were not included in this preliminary Monte Carlo program, which considered only pions. Later kaons, pseudoscalar resonances and vector resonances, as well as the radiative corrections were progressively introduced by PLUTO according to the increased need of precision.

# 5    The ϒ(9.46) and the PLUTO discovery of the ϒ→3-gluon decay

## 5.1  The ϒ Resonance

The search for the ϒ resonance at DORIS was made by scanning the center of mass energy range from 9.35 GeV upwards: initially by the PLUTO and DASP2 experiments in April - May 1978 and later by the DHHM experiment. The first results [18] were obtained by requesting at least two charged tracks and at least 2 GeV in the shower detectors. The very narrow resonance was found at 9.46±0.01 GeV (in 1977 at the Tevatron [16,17] the peak was at 9.5±0.2 GeV), with a width consistent with the machine energy spread of 8 MeV[5]. This indicated, as for the J/ψ, that the ϒ could not decay to lower energy states via 'normal' strong interactions and added considerable weight to the supposition that the ϒ was the ground state of a heavy quark-antiquark pair having a new flavour: "beauty" or "bottom". The large cross section implied a direct coupling to the virtual photon with the same quantum numbers $J^{PC}=1^{--}$ (parallel quark spins and no relative angular momentum, like orthopositronium [64]). It could then decay via 3 quanta of the strong force, 3 gluons, analogous to orthopositronium decay into 3 gammas [7,8]. From the measured total cross section it was possible to obtain the electronic width $\Gamma_{ee}\approx 1.3$ KeV [18,84] (a value closely predicted by duality arguments [110,111]), which implied a charge of -⅓ for the new heavy quark (from the models for quark binding in non relativistic potentials). The true total width $\Gamma_{Tot}$, on the basis of the total cross section, would be 20 to 60 times larger, but still much smaller than the machine energy resolution.

As stated, the new quark (and new flavour) was christened "bottom" (b quark, member of charge -⅓ of the pair whose partner of charge ⅔ was christened "top", only found in 1995 by the CDF and D0 experiments at Fermilab (Tevatron) [112,113]). The "excitation curve", which is the energy dependence of the cross section in the region of the resonance, is a superposition of the $q\bar{q}$ "continuum", the additional vacuum polarization effect of the new quark (the temporary materialization of the pair) and the "direct" excitation of the state. With more statistics and more corrections (subtraction of the expected $e^+e^-\to\tau^+\tau^-\to$hadrons contribution, effects of the detector, of the electronics and of the analysis selection, and using also neutrals), the resulting excitation curve became what is shown in Fig. 2 (first shown in [18] and as it is in [34,36,50]). Its shape is at first approximation Gaussian and at second an extremely narrow Breit-Wigner resonance. More precisely, due to the emission of gamma rays by the beams in the initial state, in 7% of the events (calculable by QED) the true initial energy is not the actual c.m.s. value, but somewhat smaller at random as initial state photon radiation lowers the c.m.s. energy to the resonant position. This process produces a small shift of the mass and a deformation of the resonance (visible in the right arm of its shape). Moreover the true final state could also be a non-resonating continuum contribution; the interference between the different processes with the same quantum numbers must also be (and has been) accounted for. An unfolding of the intrinsic width of the resonance with all these effects included is found by a fitting procedure which gives the continuous line in

---

[5] At fixed target experiment, mass resolution is dominated by the detector resolution, in $e^+e^-$ storage rings it is dominated by the beam energy resolution.

Fig. 2 (and similarly for the leptonic final states). The true total width $\Gamma_{Tot}$, on the basis of the total cross section, was calculated to be of the order of 40-50 KeV (54 KeV today [126]).

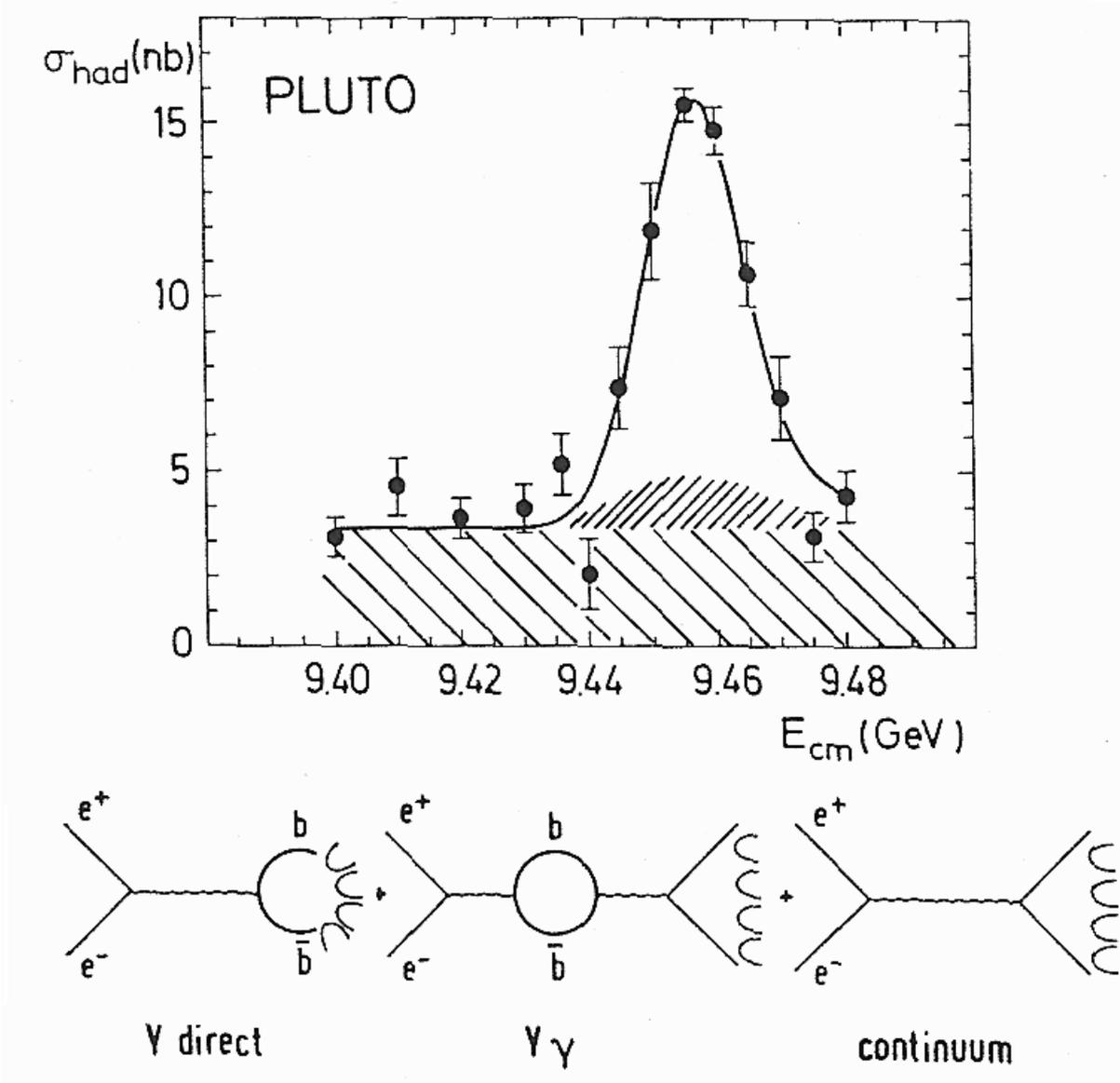

**Fig. 2.** Excitation curve of $\Upsilon(9.46)$ resonance (the $b\bar{b}$ ground state). The three contributions to the cross section $\Upsilon_{on}$ or $\Upsilon_{all}$ ($\Upsilon_{direct}$, the vacuum polarization $\Upsilon_\gamma$ and the $q\bar{q}$ continuum) are shown below and graphically distinguished by shaded areas. The solid line is a fit of the total cross section ($\tau^+\tau^-$ subtracted) with a continuous background plus a Breit-Wigner shape modified by the radiative corrections (accounting for the interference terms) and folded with the machine resolution (first shown in [18] and later in [34,36,50]). The white area ($\Upsilon_{direct}$) contains 1250 (±114) events; correspondingly below it there are 622 events [37,50].

The contemporary experiments with PLUTO at DORIS (DASP2 [19,20,92,93] and DHHM [21,22]) and those after PLUTO moved to PETRA (LENA [114], ARGUS [115-119]) together with CUSB [120] and CLEO [121-125] at CESR, found comparable results on $\Upsilon$ parameters (see Tab. 1), confirming the ones by PLUTO.[6]

---

[6] But for $<n_{ch}>$. Here the spread of the results (some not corrected) was much larger than the error bars. The different thicknesses of material in front of the detectors and the different particle identification might

| Machine | DORIS I | | | DORIS II | | CESR | |
|---|---|---|---|---|---|---|---|
| Experiment | PLUTO | DASP2[3] | DHHM[4] | LENA | ARGUS | CLEO | World average |
| Date | May 1978 | May 1978 | Sep. 1978 | 1981 | 1981/6 | 1983 | Today |
| References | [18,37,84,97] | [20,92,93] | [21,22] | [114] | [115-119] | [121-125] | PDG 2010 [126] |
| Mass [MeV] | 9456 ± 10 | 9457 ± 10 | 9460 ± 10 | --- | 9.460.3 ± 0.7 | --- | 9460.30 ± 0.26 |
| $\Gamma_{ee}$ [KeV] | 1.33 ± 0.14 | 1.35 $^{+0.11}_{-0.22}$ | 1.04 ± 0.28 | 1.23 $^{+0.10}_{-0.14}$ | 1.32 ± 0.04 | 1.30 $^{+0.05}_{-0.08}$ | 1.340 ± 0.018 |
| $B_{\mu\mu}$ [%] | 2.3 ± 1.4 | 3.2 $^{+1.3}_{-0.3}$ | 1.4 $^{+3.4}_{-1.4}$ | 3.8 $^{+1.5}_{-0.2}$ | 2.30 ± 0.23 | 2.7 ± 0.3 | 2.48 ± 0.05 |
| $B_{ee}$ [%] | 5.1 ± 3.0 | --- | --- | --- | 2.42 ± 0.14 | 2.8 ± 0.3 | 2.38 ± 0.11 |
| $\Gamma_{Tot}$ [KeV] | <180; 45 $^{+38}_{-14}$ | --- | --- | --- | 55.5 ± 4.2 | 48 ± 4 | 54.02 ± 1.25 |
| $\Gamma_{Had}\,\Gamma_{ee}/\Gamma_{Tot}$ [KeV] | 1.35 ± 0.14 | 1.23 $^{+0.08}_{-0.04}$ | 1.04±0.28 | 1.13 $^{+0.07}_{-0.11}$ | 1.23 ± 0.08 | 1.37 ± 0.06 | 1.240 ± 0.016 |
| $<n_{ch}>_{\Upsilon}$ | 8.2 ± 0.1[1] | (7.9 ± 0.7)[2] | 6.9 ± 0.6[1] | (8.1 ± 0.1)[2] | (9.1 ± 0.2)[2] | 10.17 $^{+0.05}_{-0.50}$ [1] | --- |
| $<n_{ch}>_{cont.}$ | 6.9 ± 0.1 | (6.9 ± 0.2) | 6.1 ± 0.2 | (7.2 ± 0.1) | (7.1 ± 0.2) | 8.26 $^{+0.03}_{-0.40}$ | --- |

**Table 1.** Υ data from early experiments; see PDG [126] for completeness ($n_{ch}$ missing) ( [1] $\Upsilon_{direct}$ , [2] $\Upsilon_{all}$ , [3] only particle directions, [4] no magnet, (...): uncorrected, ---: not available). Only statistical errors are considered, continuum data refer to 9.4 GeV (DORIS I), 9.98 GeV (DORIS II) and 10.49 GeV (CESR) c.m.s energies.

## 5.2  The geometry (topology): inclusive dynamics

The hadronic decays $\Upsilon_{direct}$ in Fig. 2 (the dominant decay mode by an order of magnitude) were expected to be 97-98% 3-gluon decays (with a small addition of γgg decay), resulting in average gluon energies of ≈ 4.2, 3.4 and 1.8 GeV (see Fig. 7). If gluons fragment similar to quarks certainly the fastest gluon jet can be singled out and identified safely like quark jets which start to show up clearly at jet energies of ≈ 3 GeV as had been already shown at SPEAR (Stanford) in 1975 [65,67]. Nevertheless, since the third gluon is less energetic than 3.0 GeV, the 3 gluon jets could be broad and overlapping to show up in a resolved three jet structure [77]; still they will show "flatness" and influence special topological parameters. The fastest gluon could be singled out, with a <$p_T$> a half of <$p_{||}$>; but the <$p_T$> per event (see in the following) still indicates jettiness. Still the presence of those three clusters of particles would determine [72-79] for instance a non-zero quadrupole moment for the angular distribution of the hadrons (see the event tensor in Tab. 2). The overall topology (using special variables) of each event, averaged over many or all events, could be significantly different on and off the resonance and from a random, kinematically dominated distribution (phase-space). New quantities, so called topological variables, were needed and were proposed and optimized for this purpose, chosen to be quantitatively calculable by QCD (infrared safe, except for sphericity) and based on an event tensor (see Tab. 2), derived in analogy with the inertia tensor of ordinary 3-momentum space.

---

explain the differences, due also to the non negligible presence of protons, antiprotons and charged kaons (PLUTO had to treat all charged particles as pions).

| Variables* | Definition | References |
|---|---|---|
| **Event tensor** (inertia tensor) **and $Q_k$** | $\mathbf{T_{\alpha\beta} = \sum_i (p_i^2 \delta_{\alpha\beta} - p_{i\alpha} p_{i\beta})}$ where $p_i$ are the particles' momenta; $\alpha,\beta = 1,2,3$ are the coordinate indices; the tensor eigenvalues are $\lambda_k = \{\sum_i (p_{k+1}^i)^2 + \sum_i (p_{k+2}^i)^2\}/\sum_i (p^i)^2$, ordered as $\lambda_1 \geq \lambda_2 \geq \lambda_3$; $\mathbf{Q_k = 1 - 2\,\lambda_k/(\lambda_1 + \lambda_2 + \lambda_3) = \tfrac{1}{2} \sum_i (p^i_{kL})^2 / \sum_i (p_i)^2}$ (k=1,2,3 labels the three eigenvectors)**,** $Q_1 + Q_2 + Q_3 = 1$; $Q_3 \geq Q_2 \geq Q_1$, (The optimization finds a principal axis; T=transverse, L=longitudinal to this axis; i is the index of particles in an event). | [71],[65],[32,33] |
| **Sphericity** | $\mathbf{S = 3/2\,min.(\sum_i p^2_{Ti})/(\sum_i p^2_i) = 3\,\lambda_3/(\lambda_1 + \lambda_2 + \lambda_3)}$      $0 < S < 1$ | [65],[71],[32,33] |
| **Thrust** | $\mathbf{T = max.\ \sum_i |p_{Li}|/\sum_i |p_i|}$     $\tfrac{1}{2} < T < 1$, thrust (as „spherocity", [87]) is linear and „infrared safe". | [85,86], [32,33] |
| **Triplicity** | $\mathbf{T_3 = Max(|P_1| + |P_2| + |P_3|)/\sum_i |p_i|}$ where $\mathbf{P_k = \sum_j p_{jk}}$ and k=1,2,3 is a class of particles of an event (subdivided in 3 non-empty classes) and defining 3 coplanar directions; $\theta_k$ is the angle between $\mathbf{P_{k-2}}$ and $\mathbf{P_{k-1}}$, cyclically. | [88] |
| **Acoplanarity** | $\mathbf{A = 4\ min\ (\sum_i |p_{outi}|/(\sum_i |p_i|))^2}$ where $p_{out}$ is the momentum component out of the event plane constructed by the triplicity method. | [76] |
|  | *) averaged over the events, correlated. |  |

**Table 2.** Main event topological variables

Diagonalizing the event 3-momentum tensor, gives 3 eigenvalues $\lambda_k$, the corresponding eigenvectors of which are the three principal axes of the event in momentum space. If we order these eigenvalues such that $\lambda_1 \geq \lambda_2 \geq \lambda_3$, then $\lambda_3 = \sum_i (p^i_T)^2/\sum_i (p_i)^2$ resembles the definition of sphericity and the corresponding eigenvector points in the direction of the smallest "inertia moment" (transverse relative momentum) in momentum space. $Q_k$ points in the same direction as $\lambda_k$. $Q_1$ (quadratic flatness) measures the shortest extent (smallest relative longitudinal momentum) of the event in momentum space; 2-jets events will be characterized by the smallest $Q_1$ values, 3-jet (planar) events by the second smallest and phase-space events by the largest one.

After finding the evidence for jets in $e^+e^-$ hadronic final states [32], PLUTO was able to make a systematic study of jet topological variables (variables which characterise the distribution of particles in space through a single number per event [see definitions and references in Tab. 2]) in the c.m.s. range of energies below the $\Upsilon(9.46)$ [32] and to repeat it at (and above) the $\Upsilon_{direct}$ hadronic decays (after subtraction of the continuum and of $\tau^+\tau^-$ hadronic decays) for comparison [32,33,97]. For these studies only charged hadrons were used to compute the topological variables and events with at least 4 charged hadrons were required. Uncertainties and systematic errors included the effect of the exclusion of neutrals (neutrals usually followed the direction of the fastest jet using charged particles only, but in a non-negligible fraction of events they indicated a different direction).

Two earlier examples, <S> (average sphericity) and <Q₁> (average quadratic flatness), are shown in Fig. 3 [33] as functions of the c.m.s. energy between 3.1 and 9.5 GeV and compared with the expectations of a phase-space MC, a 2-jet MC and a 3-gluon MC at the Υ energy. Below roughly 5 GeV, and including the J/ψ resonance, the topological variables were consistent with phase-space. At higher energies, <S> and <Q₁> decreased, comparing quite well with the 2-jet MC (Field-Feynman). For the Υ$_{direct}$ the data differed from both the phase-space and the 2-jet MCs, but were in reasonable agreement (see Fig. 3) with the proposed 3-gluon decay mechanism realised in the 3-gluon MC introduced by PLUTO [33], still without kaons, resonances and radiative corrections. The comparison was similar for every class of charged (and later neutral) multiplicities. This result was true as well for <sphericity>, <thrust>, <triplicity>, <Q₁>, <p$_{out}$>, <acoplanarity> and others [33] (systematic errors were not reported here). More precise results on differential distributions from a subset of the variables will be shown later (Figs. 4,5,6,9,13).

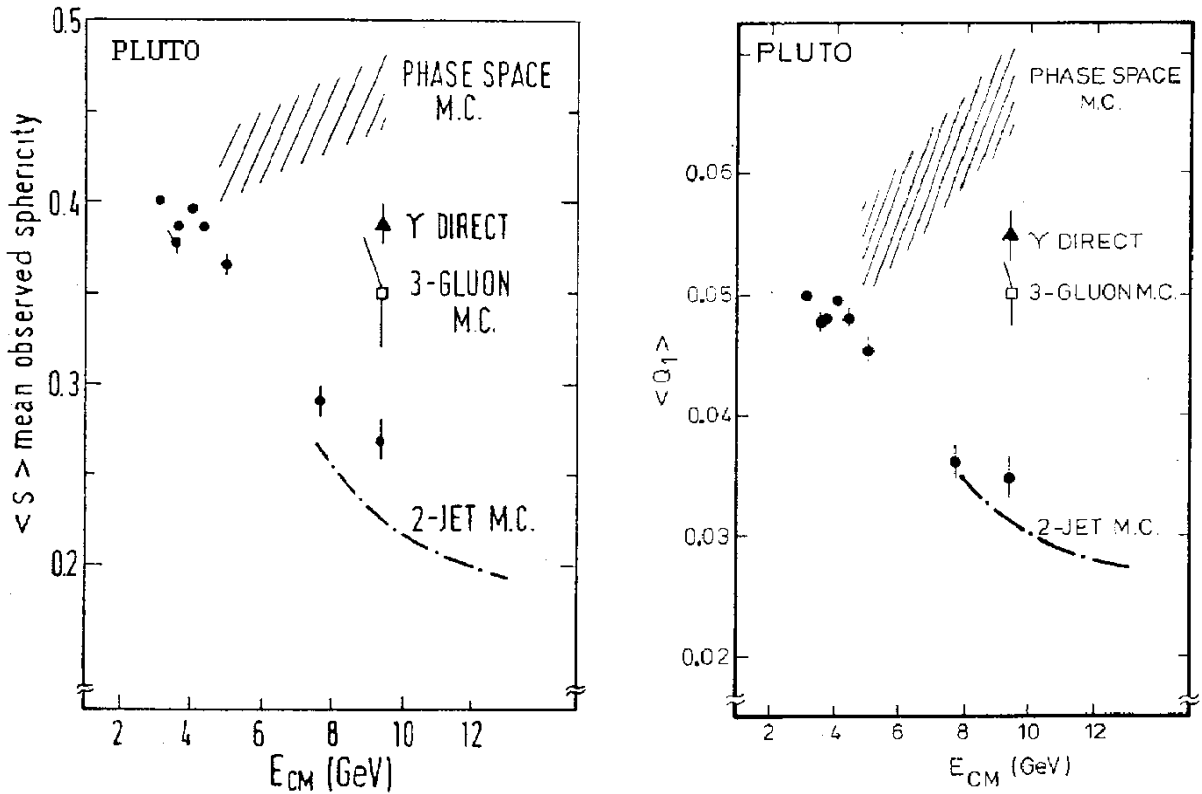

**Fig. 3.** The average observed charged sphericity <S> per event (left) and quadratic flatness <Q₁> per event (right), for events with ≥ 4 charged particles, as function of E$_{cm}$, for the continuum (black dots) and at the Υ (black triangles), compared to the expectations of phase-space MC (dashed zone), 3-gluon MC (open squares) and 2-jet MC (dot-dashed line) [33].

With the triplicity method, designed to give three directions in a plane (directions and plane fixed by kinematics in the case of 3 gluons and 3 "jets", even if not strikingly separable), 3 directions and 3 momenta (P$_i$) are determined. These momenta, as well as the corresponding relative energies x$_i$=2E$_i$/E$_{cm}$, have to correspond to the uncorrected, observed jet variables, expected by QCD at the level of gluons.

Table 3 gives a summary of the mean topological variables measured over the period Aug. 1978 − Dec. 1980 as the analysis developed: data (bold numbers, Υ$_{direct}$ and 9.4 GeV) compared with MC models. Variables used: <S> and <T> represent, transversally or linearly, how much the event is "jetty" (along a single collinear direction); <$T_3$> represents the same as <T> along three directions found by the optimization method; $x_k= 2P_k/M_\Upsilon$ is the fractional momentum along these three directions, $\theta_k$ are the angles between the three directions found by the optimization method. The process of optimization for triplicity is described in [96,97][7], the three vectors (directions) are constrained to lie in a plane and the algorithm is designed to find three and only three directions whatever the event source (including 2-jet and phase-space events). We observe two things: first, due to the different phases of the analysis, the values of the average variables did change, but were reasonably stable within their statistical and systematic error (not shown: order of magnitude less than or equal to the standard deviation of the MC numbers in the last few columns, where the systematic error was taken into account); second, in all cases the data agreed with the QCD expectation (3-gluon MC) within a standard deviation (but for <$T_3$>) and disagreed with the 2-jet FF-MC and with the phase-space MC.[8] All the results are consistent within the systematic errors mentioned above. All comparisons were decided with enough precision at June 1979 [43]: out of the 3 models, $\Upsilon_{\text{direct}}$ prefers the 3-gluon jet decay (see Tab. 3).

| | | charged | | charged + neutrals | | |
|---|---|---|---|---|---|---|
| References → | | [29,32] | [33] | [43] | [50] | [97] |
| ↓Variables | Date → | Aug. 1978 | Dec. 1978 | Jun. 1979 | Aug. 1979 | Dec. 1980 (final) |
| **<S>** | **$\Upsilon$direct** | **0.39 ± 0.02**[1] | **0.39 ± 0.01** | --- | **0.40 ± < 0.01**[1] | --- |
| | 3g-MC | --- | 0.35 ± 0.03 | --- | 0.39 ± <0.01[1] | --- |
| | PS-MC | 0.45 ± 0.02[1] | 0.46 ± 0.02 | --- | 0.49 ± < 0.01[1] | --- |
| | **9.4 GeV** | **0.27 ± 0.01**[1] | **0.27 ± 0.01** | --- | **0.28 ± 0.01**[1] | --- |
| | FF-MC | 0.25[3) 1)] | 0.22[3)] | --- | 0.28[1)] | --- |
| **<T>** | **$\Upsilon$direct** | **< 0.79** [5)] | **0.76 ± 0.01** | **0.715 ± 0.004** | **0.76 ± 0.01**[1)4)] | **0.732 ± 0.004** |
| | 3g-MC | --- | 0.76 ± 0.01 | 0.712 ± 0.003 | 0.75 ± 0.01[1)4)] | 0.72 ± 0.01[2)] |
| | PS-MC | 0.74 ± 0.02[1] | 0.73 ± 0.01 | 0.671 ± 0.003 | 0.71 ± 0.01[1)4)] | 0.69 ± 0.01[2)] |
| | **9.4 GeV** | **0.82 ± 0.01**[1] | **0.82 ± 0.01** | --- | **0.82 ± 0.01**[1)4)] | **0.808 ± 0.004** |
| | FF-MC | 0.83[1) 3)] | 0.84[3)] | --- | 0.82[1)3)4)] | 0.80 ±0.01[2)] |
| **<T₃>** | **$\Upsilon$direct** | --- | --- | **0.858 ± 0.002** | --- | **0.870 ± 0.002** |
| | 3g-MC | --- | --- | 0.850 ± 0.002 | --- | 0.86 ± 0.01[2)] |
| | PS-MC | --- | --- | 0.838 ± 0.002 | --- | 0.84 ± 0.01[2)] |
| **<X₁>** | **$\Upsilon$direct** | --- | --- | **0.855 ± 0.004** | --- | **0.862 ± 0.003** |
| | 3g-MC | --- | --- | 0.853 ± 0.003 | --- | 0.86 ± 0.01[2)] |
| | PS-MC | --- | --- | 0.819 ± 0.003 | --- | 0.83 ± 0.01[2)] |
| **<X₃>** | **$\Upsilon$direct** | --- | --- | **0.423 ± 0.006** | --- | **0.423 ± 0.005** |
| | 3g-MC | --- | --- | 0.422 ± 0.005 | --- | 0.42 ± 0.01[2)] |
| | PS-MC | --- | --- | 0.481 ± 0.004 | --- | 0.47 ± 0.01[2)] |
| **<θ₁>** | **$\Upsilon$direct** | --- | --- | **84.1 ± 1.0** | --- | **82.6 ± 0.9** |
| | 3g-MC | --- | --- | 85.5 ± 0.8 | --- | 83 ± 1[2)] |
| | PS-MC | --- | --- | 93.2 ± 0.6 | --- | 90 ± 1[2)] |
| **<θ₃>** | **$\Upsilon$direct** | --- | --- | **150 .3 ± 1.0** | --- | **151.0 ± 0.5** |
| | 3g-MC | --- | --- | 150.2 ± 0.5 | --- | 151 ± 1[2)] |
| | PS-MC | --- | --- | 144.0 ± 0.4 | --- | 146 ± 1[2)] |

**Table 3.** PLUTO average values of some topology variables at $\Upsilon$direct and at 9.4 GeV and their evolution as a function of time (publication), and comparison with the expectations (also evolving) of different models ([1)] Read from Figures with average values, [2)] Here systematic errors are included in the MC error values, [3)] No error value, [4)] converted from <1-T> to <T>, [5)] no $\Upsilon$direct available (only $\Upsilon$on), ---: not available).

---

[7] This thesis [96] was mostly done in 1978/79, first partly published in [33] and in detail only later in [97].

The 3-gluon MC describes the average values of all variables reported here (and also the ones not reported for lack of space: $x_2$, $\theta_2$, $Q_1$, $Q_2$, $Q_3$, $p_{out}$, $Q_1/Q_2$, $p_{out}/p_{in}$, acoplanarity, aplanarity ); not a single one is described by the phase-space or 2 -jet MC (as stated for instance in [36], March 1979: "the average $p_{out}$ value showed clear difference of more than three standard deviations").

Sphericity and thrust were also measured (in 1979) by the DHHM experiment [22] and found to be consistent with the PLUTO average values.

In order to improve the knowledge of topological variables, results of the differential distribution of thrust and triplicity (displayed in Figs. 4a,b), normalised energies $x_1$ and $x_3$ ($x_k = 2E_k/E_{cm}$, in decreasing order) of the three "jets" reconstructed by triplicity, and angles between "jets" ($\theta_k$ is the angle between $P_{k-2}$ and $P_{k-1}$) were presented by S. Brandt (PLUTO) at Geneva [43] and are displayed in Fig. 4a and Fig. 9.

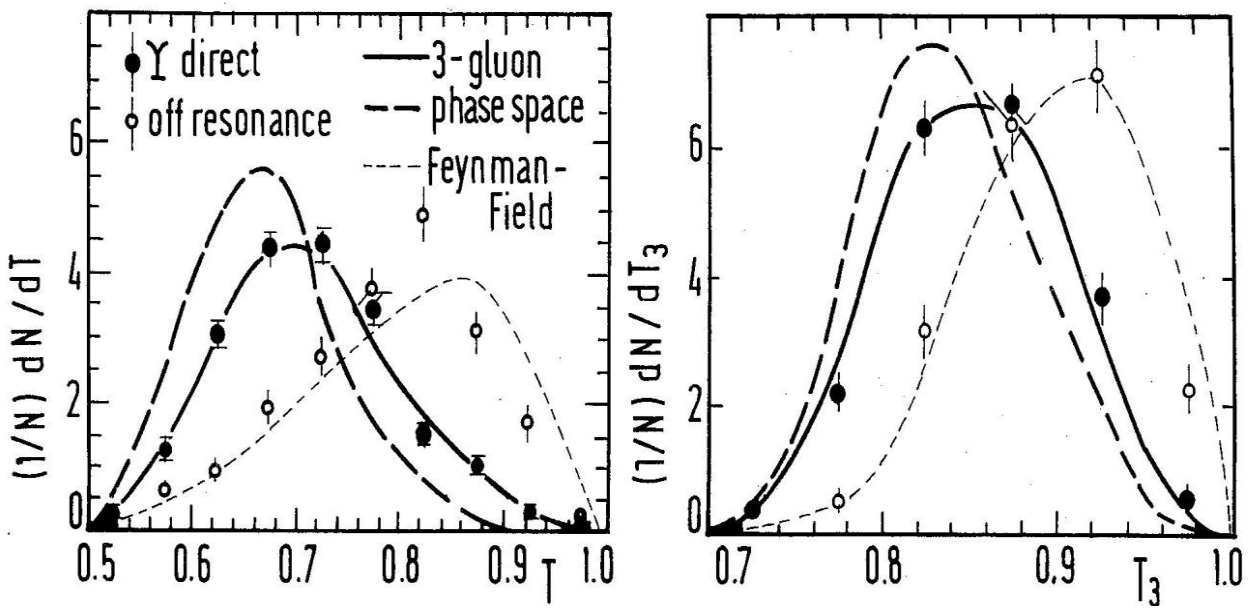

**Fig. 4a.** Experimental distributions in the variables thrust T (left) and triplicity $T_3$ (right) compared to MC calculations based on three alternative models. Charged and neutral particles were used by PLUTO to measure these inclusive variables (per event) [43]. In both cases the data agree quantitatively with the expected distributions (solid lines) for the 3-gluon-jet hypothesis (including the spin 1 hypothesis for gluons and their hadronization like quarks).

In Fig. 4a the results at 9.4 GeV (off resonance) and at $\Upsilon_{\text{direct}}$ (using now both charged and neutral particles) are compared with the differential distributions expected by 2-jet MC (F.F.), by phase-space MC and by $\Upsilon \rightarrow$3-gluon MC, three reasonable models to describe the data. About 0.06 GeV only below the $\Upsilon$(9.46) the distributions are very different (white dots) from the on resonance data (black dots) and are satisfactorily described by the 2-jet MC (F.F.). [9]

---

[8] It must be noted here that both the 3-gluon MC and the phase-space MC improved with time in terms of precision (mostly) and of statistics. The data improved with statistics, with the knowledge of the detector details and behaviour (also implemented in MCs) and with the additional use of neutral particles.

[9] The 2-jet MC (F.F.) seems here to slightly differ from the 9.4 data in case of thrust. The shift to higher values of the MC with respect to the data is probably due to the missing radiative corrections in the MC: which, for thrust close to the maximum value of 1.0 , would have the effect of a shift towards slightly lower

On $\Upsilon$ ($\Upsilon_{direct}$) the black dots are nicely described, for both the two (correlated) variables, by the 3-gluon MC and are not described (not a single point) by the phase-space MC. (The resonance distributions are just a broadening, due to hadronization and to detector resolution, of the orthopositronium matrix element [7,8,64]). On the other hand, the phase-space and the 2-jet MC behave differently by a large number of standard deviations.

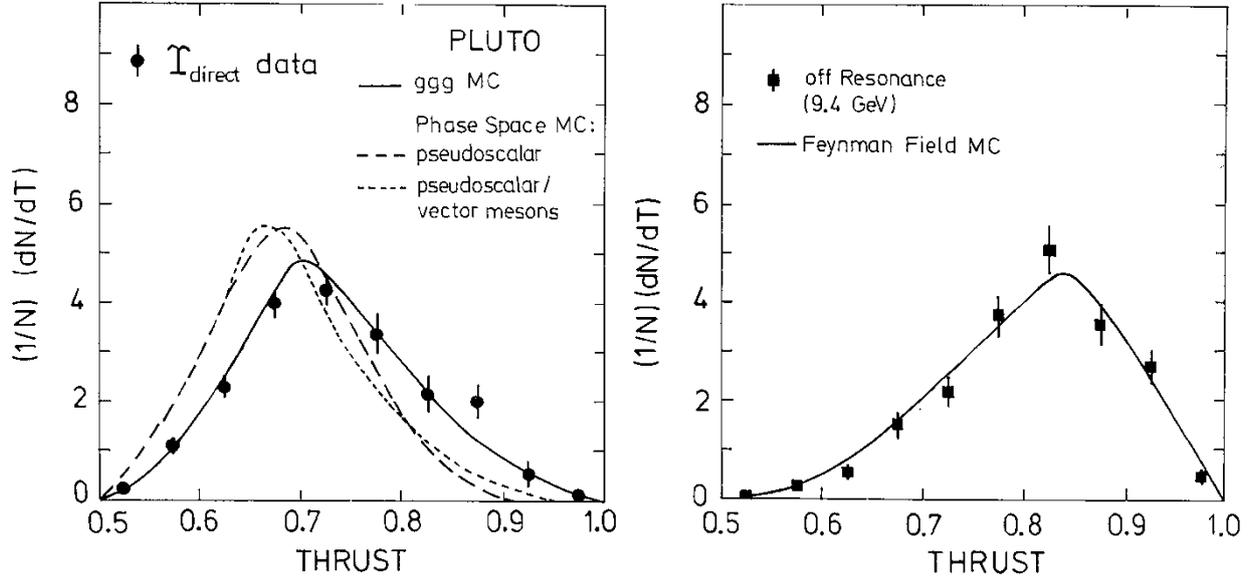

**Fig. 4b.** Experimental distributions in the variable thrust T for $\Upsilon_{direct}$ and continuum (off resonance) data compared to MC calculations based on three alternative models. Charged and neutral particles were used by PLUTO to measure these inclusive variables (per event) [97] (compared to Fig. 4a the analysis of data and models had improved).

In Fig. 4b we show in the left frame the final PLUTO thrust distribution for $\Upsilon_{direct}$ [97], compared with the final 3-gluon MC (solid line): they show perfect agreement, contrary to the new phase-space models (dashed lines), now including kaons, pseudoscalar and also vector mesons. Compared to Fig. 4a now also the continuum data (right frame, off resonance 9.4 GeV) are well described by the improved 2-jet MC (Feynman-Field).

## 5.3 Exclusion of alternative models

Another interesting topological variable is $p_{out}$, the transverse momentum out of the event plane. For 2-jet and a true 3-jet topology, this variable is non-zero due to transverse momentum broadening during the hadronization of the original partons and of course to the number of jets: for 2 jets it must be softer than for 3 jets. From the definition of a jet (limited $p_T$ with respect to $p_{||}$) the distribution of $p_{out}$ for jet events must also be narrower than for phase-space events, where there is no limitation but kinematics (assuming the same average multiplicity for all event classes). The experimental differential distribution of $p_{out}$ is shown in Fig. 5.

The experimental results (first shown in Tokyo 1978 [29]) are represented by the black dots and are again compared with the 3-gluon MC, with the 2-jet MC (F.F.) and with the phase-space values. The next Fig. 4b , data off-Y (right) and 2-jet MC (F.F.), does indeed show much better agreement with the PLUTO measured thrust distributions, after having included in MC neutrals, radiative corrections and some more details.

MC. Not only are the qualitative features listed before clearly reproduced by the data, but the Υ$_{direct}$ distribution agrees quantitatively with QCD represented by the former PLUTO 3-gluon MC [described in Chapter 4 above].

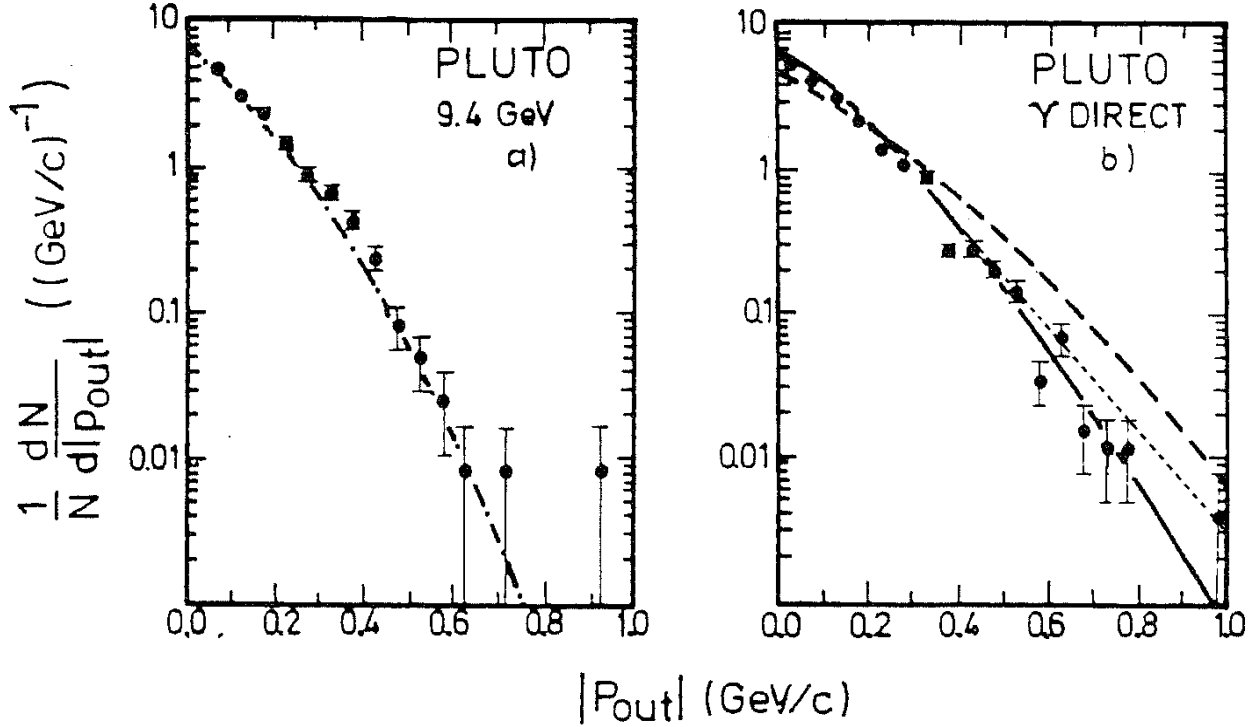

**Fig. 5.** (Reproduced from [33]) Differential distributions of p$_{out}$, the transverse momentum out of the event plane for data (black dots), **a)** at 9.4 GeV, **b)** at Υ$_{direct}$, compared to the expectations of 2-jet Field-Feynmann model (dashed-dotted line), phase-space (large dashed line) and 3-gluon decay mechanism (solid line) [29,33,36]. The short dashed line corresponds to a 'mixed model', in proportion of 50% phase-space and 50% 2-jet Field-Feynmann model [33,97] (see also Fig. 6 for thrust and θ$_3$).

Donnachie and Landshoff hypothesized [127] that the Υ$_{direct}$ data could, in principle, be described by a mixture of two models: 2-jet events and a phase-space like final state. PLUTO had considered this possibility with a 50%:50% mixture [33,97]. The result is shown in Fig. 5 (short dashed line) and Fig. 6 where thrust and θ$_3$ (the angle between "jet" 1 and 2, see triplicity in Tab. 2) differential distributions are compared to the 3-gluon MC (solid line) and to the 'mixed model' (short dashed line). Although the mixed model gets closer to the data than the two separately rejected models shown in Fig. 4a, in both cases the data are clearly better described by the 3-gluon model (see also Fig. 6). (Apart from this fact there is no plausible theory behind the Donnachie-Landshoff proposal.)

A complete quantitative test of the mixed model would allow the percentage of the two components to be determined by a fit to all topological variables simultaneously. However, as the variables can be strongly correlated, a simpler approach is to consider just three independent variables: thrust, θ$_3$ and p$_{out}$.

The best percentage of the phase-space and 2-jet MC models has been determined from the three independent variables individually with the results: for p$_{out}$, about 30% phase-space and 70% 2-jet model; for thrust and θ$_3$, an almost opposite 70% phase-space and 30% 2-jet. The combined fit to these three variables gives an intermediate mixture of 53%:47% (phase-space:2-jet), broadly compatible with the 50%:50% one displayed in Fig. 5b and 6. The χ$^2$/n$_{dof}$ value (n$_{dof}$ is the number of degrees of freedom, here $n_{dof}$=32) for the mixed model fit is 4.2 to be compared to the $\chi^2/n_{dof}$ of 1.1 for the 3-gluon MC.[10]

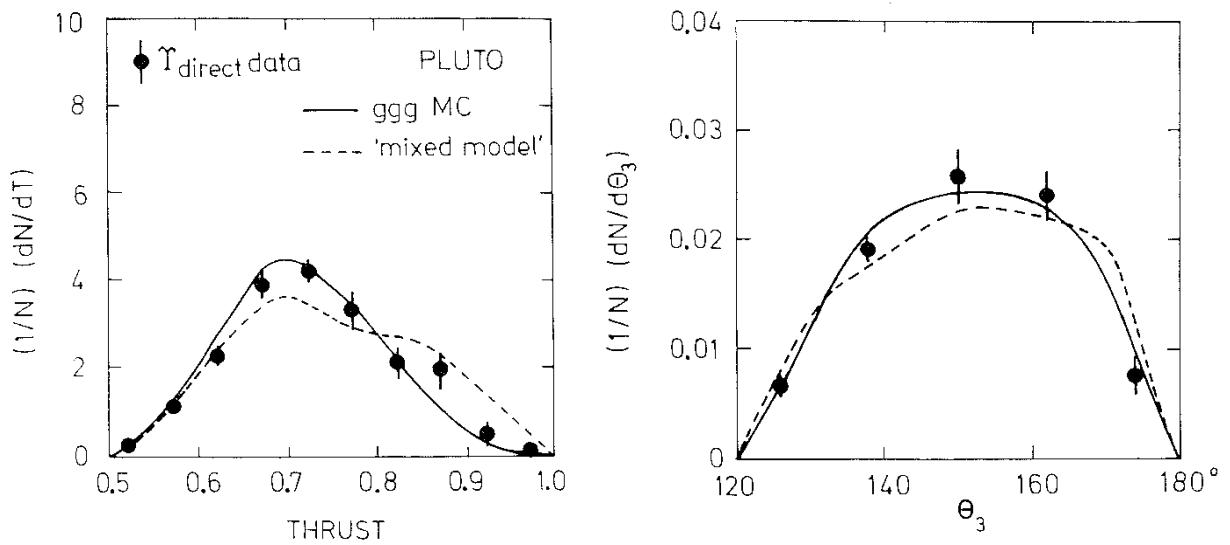

**Fig. 6.** Thrust and $\theta_3$ distributions of $\Upsilon_{direct}$ data (black dots) compared to the expectations of 3-gluon MC (solid line) and a 'mixed model' of phase-space MC and 2-jet MC (dashed line) in proportion of 50%:50% [33,97] (see also Fig. 5b for $p_{out}$).

We conclude that the topology of the full $\Upsilon_{direct}$ events, studied by PLUTO using many variables, some not shown here (details in the "Topology" paper [97] and in earlier PLUTO papers) agreed fully with the 3-gluon MC and disagreed with the continuum data, the 2-jet MC, with all versions of the phase-space MC as well as with an arbitrary mixture of the phase-space MC and the 2-jet FF MC.

A colourless 3-gluon model was also excluded by the PLUTO data, because it behaves like the 2-jet F.F. model [128], already excluded by the PLUTO data.

These results give the logical *necessary* conditions for the demonstration of the QCD 3-gluon-jet hypothesis: no other reasonable hypothesis described the data better or even at the same level as the proposed 3 gluons of spin-parity $1^-$. All other reasonable alternative hypotheses [2-jets; 3 scalar gluons; 3 colourless $1^-$ gluons; phase-space models (with only pions, with kaons, with additional pseudo-scalar resonances, with additional vector mesons); a fitted mixture of 2-jet model and phase-space] gave *substantially worse* descriptions of the data.

## 5.4 The exclusive 3-gluon dynamics

The kinematics of the $\Upsilon \rightarrow$3-gluon decay is fixed by the 3 gluon momenta $P^g_k$, or by the 3 gluon energies, and by the angles $\theta_k$ between the three vectors. The scaled momenta of modulus $x^g_k$=2 $P^g_k/M_\Upsilon \sim P^g_k/E_{beam}$ have vector sum zero and scalar sum 2. The 3 gluon vectors $P^g_k$ (estimated by the three triplicity momenta $P_k$: see later Fig. 8) might be experimentally identified in every event by their scaled momenta, ordered as $x_1 \geq x_2 \geq x_3$ (these variables are more flexible, since they can be measured also by subsets of momenta, e.g. charged-particle momenta). For massless gluons, the following relationship holds: $x_k$= 2 $\sin\theta_k/\Sigma_j\sin\theta_j$.

---

[10] A few years later CLEO, using thrust only, found similar best mixture values. ARGUS found an upper limit of 5% for the 2-jet component on $\Upsilon_{direct}$, using only thrust (see Fig. 13).

The dynamics of the Υ→3-gluon decay (for $1^-$ gluons) is the same as for the orthopositronium $3\gamma$ decay [64,7,8]. The matrix element in leading order QCD gives the momentum distribution of the gluons [9-11,74]:

$$\frac{1}{\sigma}\frac{d\sigma}{dx_1^2 dx_2^2} = \frac{6}{(\pi^2-9)}\{x_1^2(1-x_1)^2 + x_2^2(1-x_2)^2 + x_3^2(1-x_3)^2\}/x_1^2 x_2^2 x_3^2 \ .$$

For scalar gluons, the momentum distribution would be [128]:

$$\frac{1}{\sigma}\frac{d\sigma}{dx_1^2 dx_2^2} = \sim\{ (x_1-x_2)(x_3-x_2)(x_1+x_3-x_2) + \text{even permutations of 1,2,3}\}/x_1^2 x_2^2 x_3^2 \ .$$

Since the scaled momenta $x_k$ and the angles between gluons $\theta_k$ are functionally related, the above formulas can be transformed into angular distributions.

The density distribution of final states is obviously different in the two cases and has been studied theoretically by Walsh and Zerwas [128]. We show it in Fig. 7.

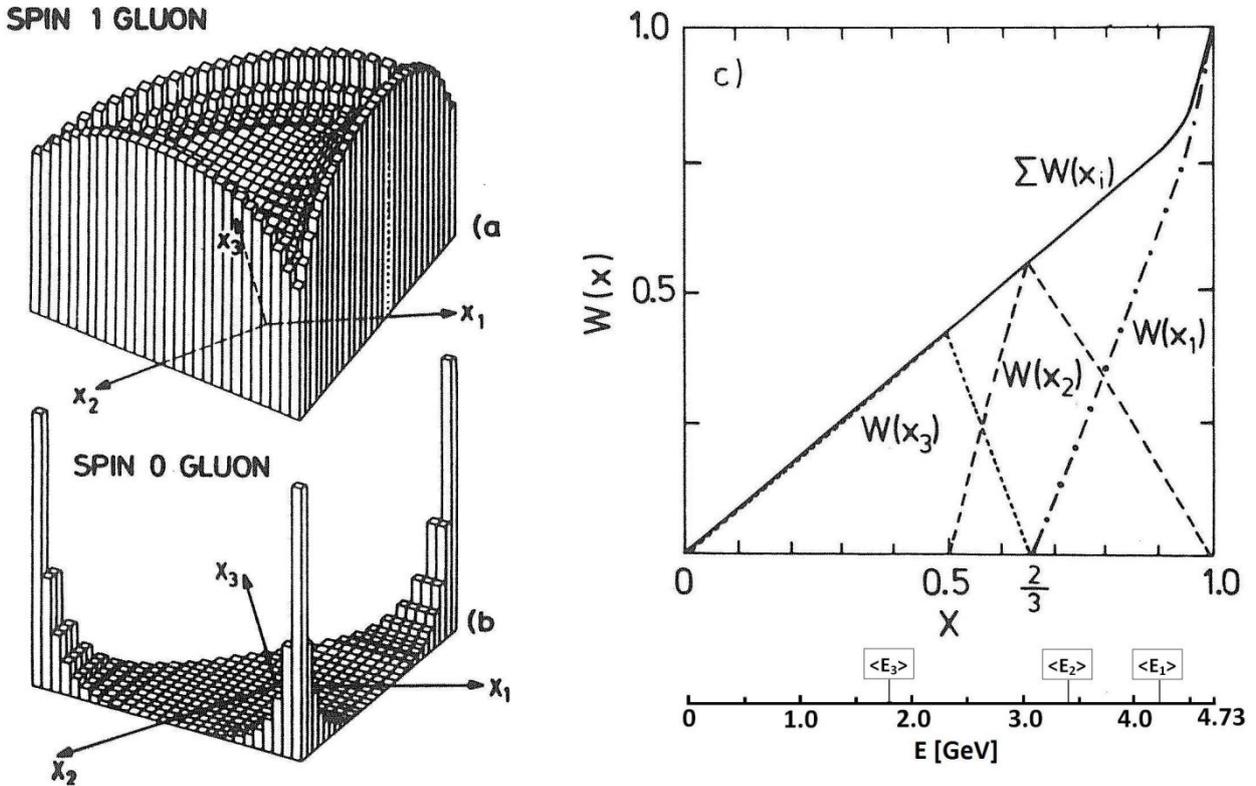

**Fig 7.** Density distribution of 3-gluon decay for a) vector gluons and b) scalar gluons; c) projections of the vector gluon distributions onto the axis of the Dalitz plot [128]. The 3 vector gluons have average energies of $<E_1>$=4.2, $<E_2>$=3.4, $<E_3>$=1.8 GeV (see added energy scale for the Υ case) and the fastest one has enough energy to produce a visible jet (as at 8.4 GeV c.m.s. energy in $e^+e^-$ annihilations).

As we can see, in case of the scalar gluon the distributions of the fractional energies are peaked at the two extremes; as a consequence, the most frequent event would be the configuration with two gluons of relatively large momentum with the third gluon of low momentum, almost a two-jet configuration, already excluded by the topological variables.

Up to this point, we have tested the expected gluon kinematics and dynamics in Υ hadronic decays (3 jets with the expected distribution of angles in between and the corresponding distribution of fractional momenta) only inclusively, by the thrust, triplicity and $p_{out}$ (as well as $\theta_3$) distributions. Was the Υ mass large enough and the PLUTO detector and its methods good enough for exclusive tests of QCD gluon dynamics using event-by-event reconstructed 3-gluon angles and energies? These important questions were studied using MC simulations in the PhD thesis [96] of one of us (H.-J. Meyer) . The answer is shown in Fig. 8.

The 3-gluon MC was a full and tested simulation of the PLUTO experiment. For every "detected" and "reconstructed" event, after computing the triplicity 3-vectors (which lie in a plane; see definitions in Tab. 2) by using the measured momentum and energy of charged and neutral particles and having ordered the vectors in decreasing order of energy, the directions (as well as the angles $\theta_k$ between them) are defined. We can compare those final state directions with the directions of the generated 3 gluons (ordered independently in the same way). In the ideal case (massless particles, perfect detection, including all generated particles, and perfect reconstruction), the angle $\delta$ between them should peak at zero by energy and momentum conservation. A Gaussian spread around zero ($|\cos \delta|=1.0$) is expected (if the method has small systematic errors).

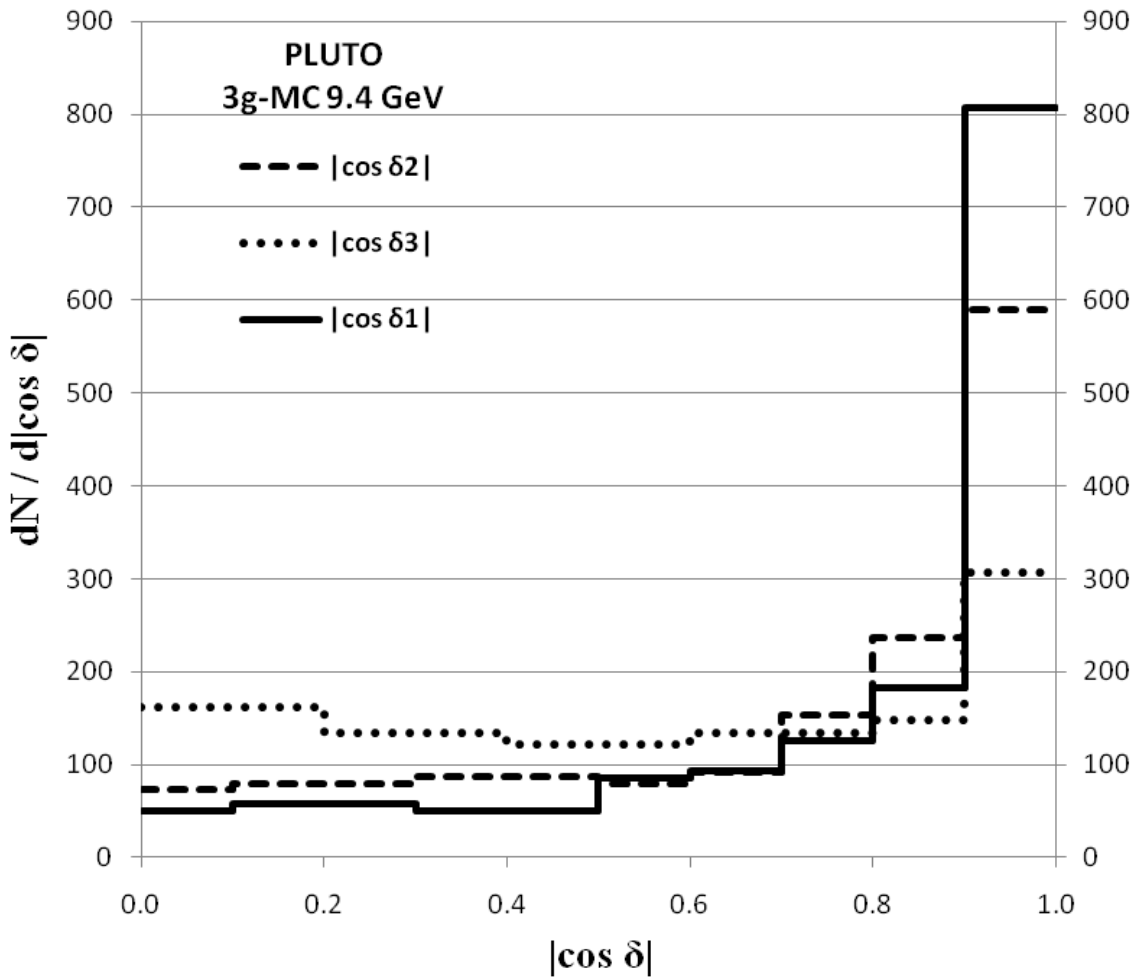

**Fig. 8.** Comparison between the generated gluon and the reconstructed jet axes using the triplicity method, ordered separately from the fastest to the slowest gluon and jet. The distributions of the absolute cosine of the angle $\delta_i$ between the corresponding jet and gluon directions are shown. The most frequently reconstructed jet direction is the correct one to within the bin $|\cos\delta|=0.9$-1.0. (Reproduced from [96])

*Even if the three jets were not easily identifiable individually by the sphericity as at SPEAR or thrust as at DORIS (but in principle for the fastest jet with $<E_1>$=4.2 GeV), their best directions were*

*found and with results comparable to those from the 3-gluon MC [29,43].* This was the main test of the matrix element.

Even if the jet energies match less well than the directions to their gluon equivalents, still, at reconstructed level, the measured energy inclusive variables compare well with the MC expectations. This result is shown in Fig. 9, where the distributions of $x_1$ and $x_3$ (fractional observed energies of the first and the third triplicity jet) and the corresponding $\theta_1$ and $\theta_3$ angles are displayed for the $\Upsilon_{direct}$ (black points) and the off-resonance continuum (open points). The data are compared with the appropriate physics models (3-gluon MC, continuous line, and Field-Feynman 2-jet MC, short dashed line, respectively) as well as with the phase-space MC (dashed line).

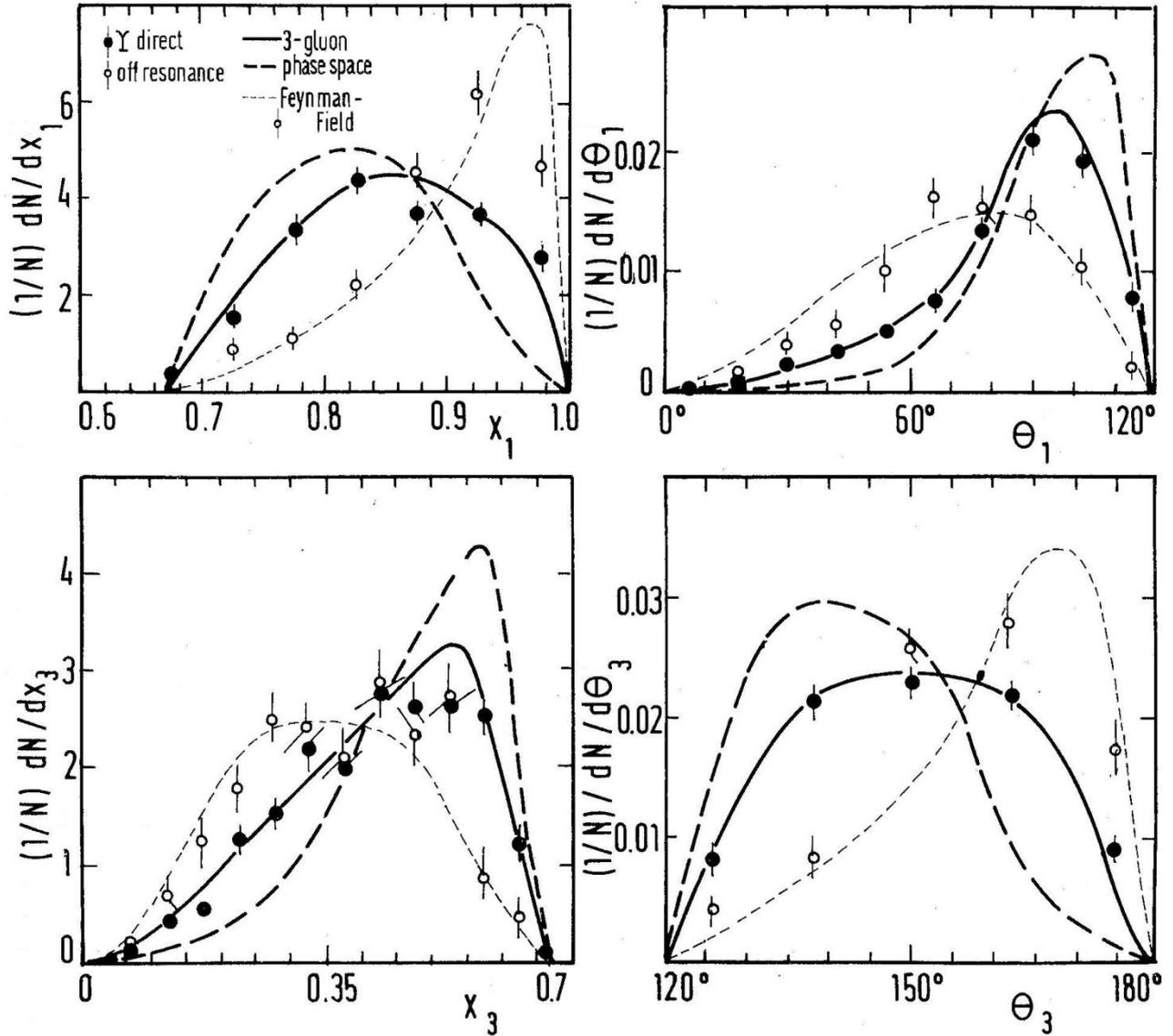

**Fig. 9.** Experimental distributions [43] of inclusive variables. Reconstructed fractional gluon energies $x_1$, $x_3$ and reconstructed angles $\theta_1$, $\theta_3$ between gluon jets compared to Monte Carlo expectations. Charged and neutral particles were used by PLUTO to measure these inclusive variables (per event). Labels as in Fig. 4: full points ($\Upsilon_{direct}$), open points (off resonance), solid line (3-gluon MC), dashed line (phase-space MC), dotted line (Field-Feynman 2-jet MC). Note the different scales.

The good agreement of the $\Upsilon_{direct}$ decay points in Fig. 9 with the 3-gluon MC means that the perturbative QCD expectation for the 3 partons *survives hadronization* and is strictly confirmed by the data. (Of course, this result is valid also for the $x_2$ scaled energy of the second intermediate jet and for $\theta_2$, both not shown here).  In average, PLUTO got $<E_1>$=4.08±0.01, $<E_2>$=3.38±0.02 and $<E_3>$=2.00±0.02 GeV [97], strictly comparable with the theoretical values indicated in the captions of Fig. 7.

The QCD expectation implies also the quantum numbers, here the production of three $1^-$ gluons with the correct angular distributions (matrix element, [7,8,75,128,129]). The angular distribution of the sphericity axis on $\Upsilon_{direct}$ (here constructed with only charged particles and strongly correlated with the thrust axis) of the fastest jet with respect to the $e^+$ beam is shown in Fig. 10a [33] and 10b [97], compared with the expected angular distribution according to the three-vector-gluons MC.  Fig. 10a shows the early PLUTO result (December 1978), as observed and without subtraction of the $\tau^+\tau^- \rightarrow$ hadrons contribution, compared with the expected spin 1 distribution of the 3 gluons under the same conditions (solid line). The agreement is impressive. A few months later Koller and Krasemann [129] compared these results with their expectation for 3 scalar gluons, the PLUTO data rejected this hypothesis. Fig. 10b [97] shows the data (here for the thrust axis) now corrected for detector and method effects and $\tau$ subtracted, compared with curves for vector (solid line) and scalar gluons (dashed line). The level of agreement is less pronounced, but the data again prefer the spin 1 hypothesis. The best fit gives $1+(0.83\pm0.23)|\cos\theta|$, compatible with $\sim 1+0.39 \cos\theta$ ([129] for spin-1), and disagreeing with $1-0.995|\cos\theta|$ of the scalar gluon hypothesis. Later LENA [114] found 0.7±0.3 and CLEO [122] 0.32±0.11 for the coefficient to $|\cos\theta|$, all compatible with 0.39.

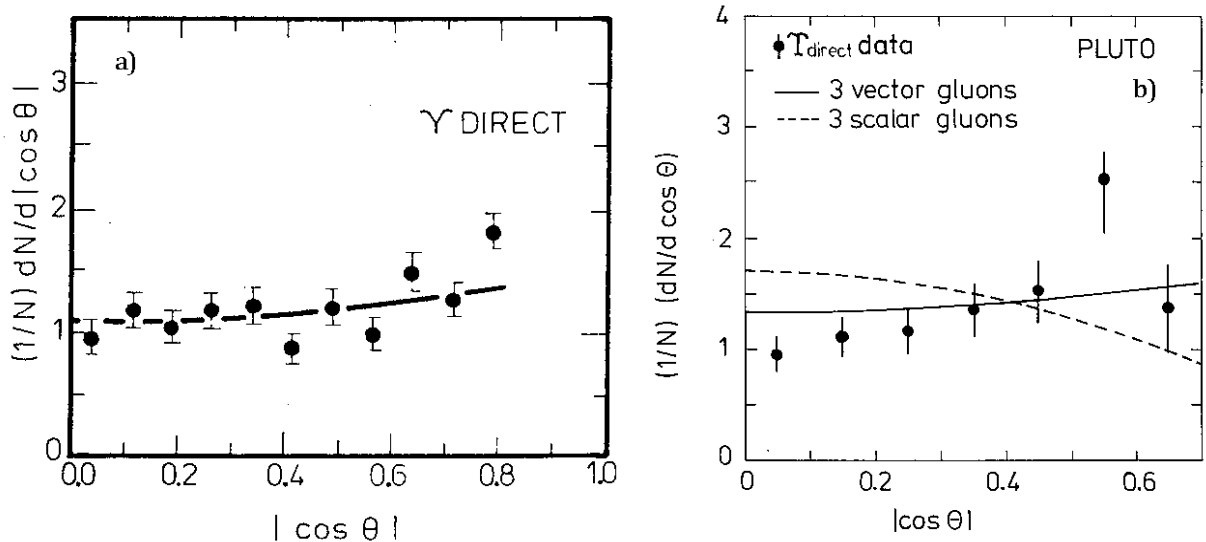

**Fig. 10a.** Observed PLUTO angular distribution of the sphericity axis (for charged particles only) with respect to the positron beam axis, solid line: QCD expectation [33]. **10b.** Corrected and $\tau$ subtracted angular distribution of the thrust axis (for charged and neutral particles) [97].

Since the spin-½ hypothesis would imply two partons instead of three and the 2-jet hypothesis was already excluded by topology and since it is not possible to make spin 1 (as the virtual photon) with three partons of spin 2, once spin 0 was excluded by PLUTO only the 3 spin-1 gluon hypothesis remained.

In his talk at Geneva 1979 on *Jet analysis* [45], P. Söding (TASSO) said on page 272: "*The PLUTO data are in very good agreement with a three gluon decay model*" and at page 273: "*In spite of the low energy there's a clear distinction, both in the data and the model, from simple phase-space like behaviour*".

The expected 3-gluon topology and the matrix element [43] as well as the gluon's quantum number [33] were checked in the first half of 1979. A single theory (QCD) was able to describe in great detail all the macroscopic features of the ϒ hadronic decays.

## 5.5 The hadronization of the gluon

To measure the ϒ → 3 gluons → 3 jets hadronization corresponds to measure the gluon→jet hadronization at three average energies: the fastest jet has 4.2 GeV average energy, like for jets at 8.4 GeV c.m.s. energy, were jets were clearly singled out; the other two have 3.4 and 1.8 GeV. In this sense, the inclusive particle distribution at ϒ, using ϒ→3-gluon decay dynamics, can be interpreted as inclusive particle distribution of the gluon→jet hadronization. Even if, as in PLUTO, particle identification is limited, only for gammas, electrons and $K^0_s$'s, and if pions and kaons are treated as the same particle, the charged and neutral particle multiplicities can be measured. Also the $K^0_s$ production was measured.

The PLUTO 3-gluon MC agreed both with the inclusive features of the ϒ decays (for both topological variables and inclusive variables), and the main aspects of the approximated hadronization model in the MC were confirmed. The more detailed fragmentation properties (particle multiplicities, momentum distributions, flavour blindness, two particle correlations [133]) were also explored and seen to be qualitatively (due to the low statistics and low precision) closer to the expectation for gluon jets rather than quark jets. The results were presented and discussed by H. Meyer (PLUTO) at Batavia [50] and in [97].

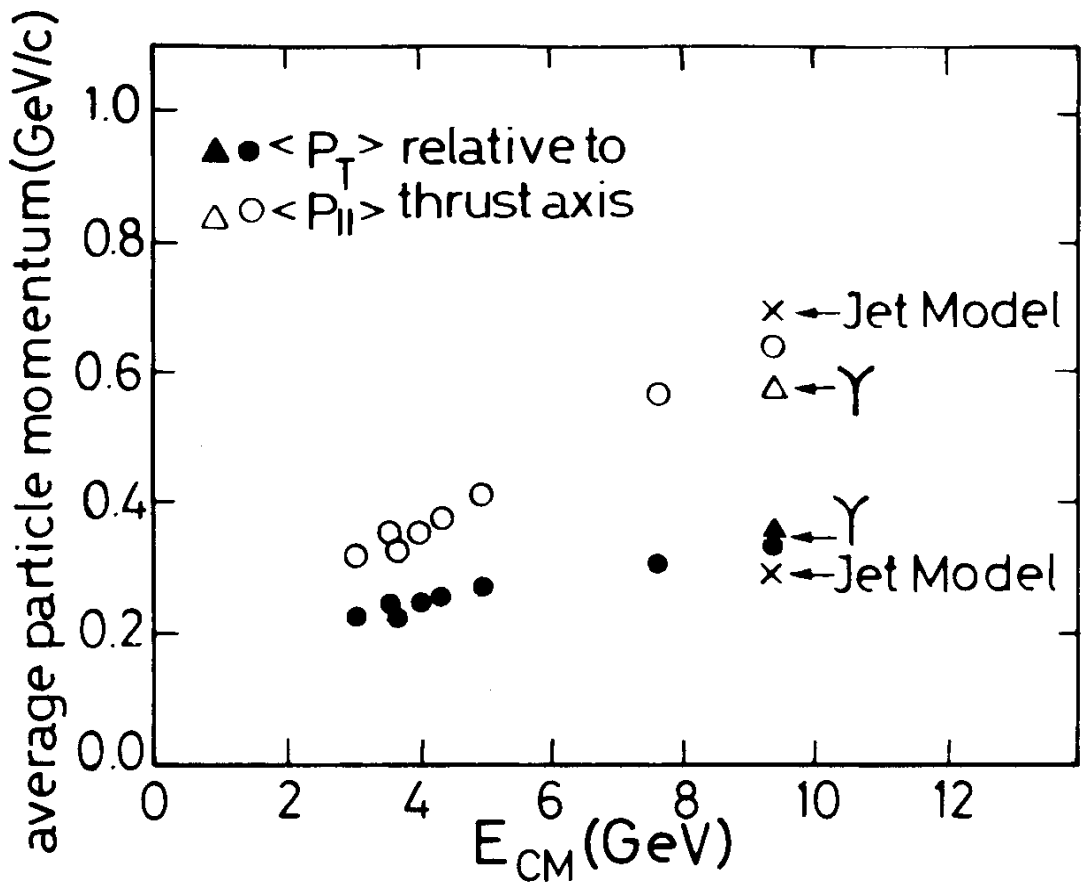

**Fig. 11.** Observed event $<p_T>$ and $<p_{||}>$ with respect to the thrust axis as a function of c.m.s. energy. The open triangles show the result of ϒ data ("ϒ$_{on}$") and the crosses the result of the 2-jet MC (F.F.) [32].

Concerning the charged particle multiplicities, as they are shown in Tab. 1 (last two lines: $\Upsilon_{direct}$: 8.2±0.1, continuum: 6.9±0.1), the multiplicity on $\Upsilon$ is larger than in the continuum by 1.3±0.2 units. If the quark fragmentation is assumed for the gluons, by extrapolation to the same energies, a multiplicity of 7.5±0.3 [35,97] would be expected, smaller than 8.2±0.1 found. On the other hand, this PLUTO experimental multiplicity on $\Upsilon$ is a bit smaller than the expected asymptotic gluon fragmentation, predicted to produce order of 9 charged particles at $\Upsilon$ [73][11]. These PLUTO results (heard at 1978 Copenhagen Conference [24]) encouraged G. Gustafson [139] to go ahead with the string fragmentation model [104-107], a possible explanation of the remaining difference. Today this outcome looks as another important consequence of PLUTO results.

Concerning the average momenta, in Fig. 11 ($\Upsilon_{on}$) and in Tab. 4 ($\Upsilon_{direct}$) we display the average inclusive PLUTO results [32,33,37,41], not corrected for detector and event selection effects (the MC values were partly not the final ones; only statistical errors were included). In addition, for the slope B {dσ/dx = A exp(-Bx)}, corrected results from CLEO are given [125]. The momenta are related to the thrust axis ($<p_{||}>$, $<p_T>$) and to the event plane ($<p_{out}>$).

| Data type | PLUTO (observed) | | | | CLEO (corr.) |
|---|---|---|---|---|---|
| | $<p_{||}>$ [GeV] | $<p_T>$ [GeV] | $<p_{out}>$ [GeV] | Slope B | Slope B |
| $\Upsilon_{direct}$ | 0.49 ± 0.01 | 0.34 ± 0.01 | 0.129 ± 0.003 | 10.9 ± 0.3 | 11.6 |
| 3g-MC | 0.55 ± 0.01 | 0.38 ± 0.01 | 0.140 ± 0.006 | 8.9 ± 0.2 | --- |
| PS-MC | 0.58 ± 0.01 | 0.47 ± 0.01 | 0.177 ± 0.006 | 10.7 ± 0.1 | --- |
| continuum | 0.62 ± 0.02 | 0.33 ± 0.01 | 0.118 ± 0.003 | 7.8 ± 0.2 | 8.9 |
| FF-MC | 0.72 ± 0.01 | 0.32 ± 0.01 | 0.115 ± 0.002 | 7.8 ± 0.1 | --- |

**Table 4.** Event average values of momentum components and momentum distribution slopes for $\Upsilon_{direct}$ and continuum data (at 9.4 GeV for PLUTO and at 10.49 GeV for CLEO) compared to model predictions (MC) and CLEO results (---: not available).

In Fig. 11 the $<p_T>$ at $\Upsilon$ is approximately the same as for the sequence of values found for lower energy quark jets; $<p_{||}>$ is smaller, as expected for three non collinear jets instead of two. In Tab. 4, the most striking feature in the PLUTO results is that the event $<p_T>$ at $\Upsilon$ (0.34±0.01) is strictly comparable with the one for 2-quark jets at 9.4 GeV (0.33±0.01) and much smaller than 0.47±0.01 for the PS-MC. Already by comparing the uncorrected PLUTO data results delivers a model independent strong indication for jettiness: kind of two jet formation. The hypothesized jets would be broader (as indicated by $<p_T>/<p_{||}>$) than quark jets at the same energy. But the fastest jet data in a hemisphere of the event (not available here) would show even larger $<p_{||}>$ (≈0.61 GeV) by linearity of it with energy with almost the same $<p_T>$ (≈0.34 GeV) (see Fig. 11).[12]

---

[11] Including systematic errors would increase the corrected charged multiplicity in case of PLUTO (see footnote 6).

[12] Unfortunately important considerations of P. Söding's recent publication [138] concerning the $\Upsilon$ (but the paper concerns mostly gluon bremsstrahlung) are based on a statement (page 5) not corresponding to PLUTO measurements: "*The average particle momentum (at $\Upsilon$) is only 0.40 GeV and thus not much larger than the average transverse momentum in a typical jet.*" Instead, according to our Tab. 4, $<p_T>$ is 0.34 GeV

Whether the 9.4 GeV data rather well agree with the two quark jet model (FF-MC), it is obvious that the average charged particle properties of the ϒ in Tab. 4 are less well reproduced than the topological event quantities in Tab. 3 by the 3-gluon MC model used here (with quark-like fragmentation)[13].

The inclusive charged particle momentum spectrum is well represented by  $d\sigma/dx = A \exp(-Bx)$ (apart from the last few bins). For the slope B, the PLUTO value for the ϒ data is 10.9 (later compatible with the corrected 11.6 from CLEO), larger than the 3-gluon model expectation of 8.9. For the 9.4 GeV data the values of B for data and 2-quark-jet MC were exactly the same (both 7.8 uncorrected, CLEO: 8.9 corrected). The independent quark fragmentation MC model, as used in PLUTO's 3g-MC,  describes the ϒ data reasonably well. The slightly larger average momenta in the MC might be due to the missing radiative corrections in the simulation  (the same is true for FF-MC). Since in PLUTO's 3-gluon model the fragmentation was approximated by that of quarks, the specific effect of gluon fragmentation and hadronization was not properly accounted for. In the data it is expected to produce more low energy pions and kaons than in MC. This fact is reflected in $<p_{||}>$ and $<p_T>$: the 3g MC has in fact a larger $<p_{||}>$ (0.55) and $<p_T>$ (0.38) and a smaller slope B (8.9) compared with the  ϒ data (0.49, 0.34 and 10.9, respectively), corresponding to the slightly (≈10%) smaller multiplicity ($<p>$ is 0.60 GeV for data and 0.67 GeV for 3-gluon MC). In addition, B in ϒ data larger than in the continuum  is compatible with more than 2 jets (already supported by the same average sphericity behaviour, see Fig. 3). JADE [108], as PLUTO, found smaller $<p_{||}>$ ($\sim<n_{ch}>^{-1}$) and larger $<p_T>$ for the slowest jet at PETRA high energies (expected to be very often the gluon) compared to quark jets (2-jet events). Although for $<p_T>$ this was not evident in the PLUTO data, in the case of $<p_{||}>$ 0.49 at ϒ was definitely smaller than 0.62 in the continuum (also later found by JADE) and corresponding qualitatively to the larger expected multiplicity.

The same argument is valid for $<p_{out}>$. This linear flatness parameter is very small, as observed for 2-quark jet events (118±3 MeV) and only slightly larger for ϒ$_{direct}$ events (129±6 MeV), a result without use of models, and about one third of the event $<p_T>$, as qualitatively expected for 3 jets, being the $<p_{out}>$ correlated to the event $<p_T>$ projected orthogonally to the event plane. It is even flatter than expected by the 3-gluon MC in PLUTO (140±6 MeV): as we know the data do prefer a larger charged multiplicity (8.2) than in MC (7.5), in the direction of a more specific gluon hadronization, lowering $<p_{out}>$ exactly in the inverse proportion of multiplicities. Anyway, the experimental value with the very small $<p_{out}>$ but larger than for 2-quark jet events is at least qualitatively consistent with a flat 3-jet event, as expected by ϒ->3g->3-jet theory

In conclusion, these variables (Tab. 4 and Fig. 3), indicating jettiness ($<p_T>$), more than 2 jets (sphericity and B) and flatness ($<p_{out}>$), were already very suggestive in a model independent way for the 3-gluon→3-jet hypothesis, as confirmed later by the comparison of differential distributions of the topological variables with the models and by the further studies already shown. Of course, the studies in this Chapter on multiplicities and average momenta of ϒ hadronic decays have been the first ones on gluon hadronization (once accepted the ϒ→3-gluon interpretation), hence pionering, but not yet conclusive.

---

and $<p_{||}>$ = 0.49 GeV which means $<p>$ is 0.60 GeV and the jet $<p_T>$, especially for the fastest jet (with $<E>$=4.2 GeV), is substantially smaller than $<p_{||}>$, as expected for parton-jet hadronization (Fig. 11). The fastest jet in Y$_{direct}$ decays is kinematically similar to quark jets at  8.4 GeV c.m.s. energy, very 'typical' jets indeed, and has by kinematics even larger $<p_{||}>$ (≈0.61 GeV, as previously mentioned).

[13] It is to be noted that topological variables, such as sphericity or thrust, are more influenced by the high momentum particles (sorted first), where the "memory" of the original parton resides. The non-perturbative part (hadronization) influences more the most frequent  low energy particles.

Approximately the same number of kaons and anti-protons in $\Upsilon$ decay (3 gluon jets) as in 2-quark jet events at 9.4 GeV had been observed by DASP2 [19,20] already at the time of 1978 Tokyo Conference [28]. As far as PLUTO was concerned, we found somewhat more $K^0_s$'s in $\Upsilon$ decay than in the continuum [31,36]. Since the 3 gluon jets each had less average energy than the 2-quark jets (4.7 GeV) at the same c.m.s. energy and since the jet multiplicity increases logarithmically with energy, these experimental results on strange meson and baryon production meant probably (once demonstrated by PLUTO that $\Upsilon$ decayed through 3 gluon jets) *that the gluon hadronizes differently from quarks.*

Figure 12 [130] shows the $K^0_s$ momentum distribution, corrected for $K^0_s$ detection efficiency off resonance (left) and for $\Upsilon_{direct}$ decays (right). The $\Upsilon_{direct}$ distribution in x = 2p/√s falls faster than the continuum, for the same reason as in Tab. 4; but this time the MC includes the gluon specific fragmentation [130] and describes perfectly the data in shape and rate. Each $K^0_s$ distribution has a shape similar to the respective inclusive distribution of charged particles (parametrized in Tab. 4).

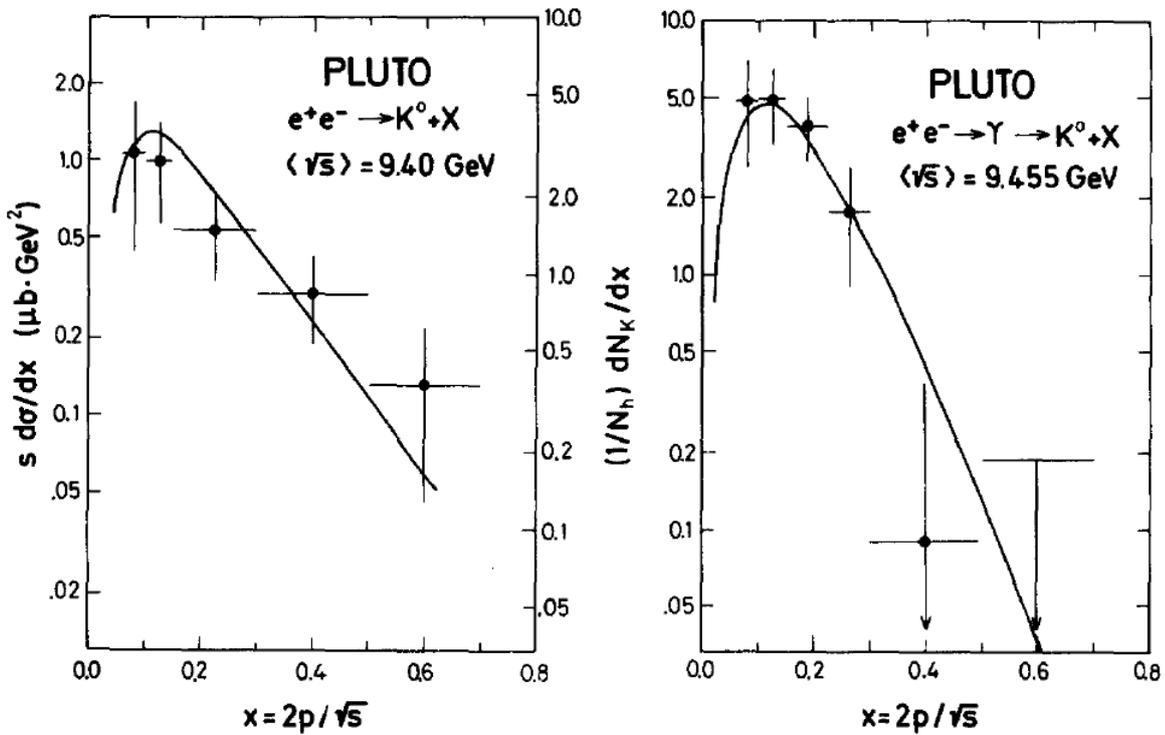

**Fig. 12.** Comparison of the corrected $K^0_s$ yield per event for continuum and $\Upsilon$ data [130]:
- left: s dσ/dx for the continuum at $E_{cm}$ = 9.4 GeV. Also given is the yield per event $1/N_h$ ($dN_K/dx$) with $N_K = 2NK^0_s$ (see right hand scale). The curve indicates the model prediction.
- right: yield per $\Upsilon$ event, $1/N_h$ ($dN_K/dx$) for the $\Upsilon_{direct}$ decays after background subtraction. The curve represents the prediction of a $\Upsilon \rightarrow$ 3-gluon decay model now with the expected gluon fragmentation according to the Lund MC.

The corrected $K^0_s$ yields which PLUTO found are given in Tab. 5 [130], where the number of $K^0_s$ per event are compared with ARGUS [118,119] and CLEO [125] results.

With much improved statistics, the ARGUS and CLEO results compare well with PLUTO. The slightly increased fraction of $K^0_s$ is confirmed. The share of $K^0_s$'s among the final state particles is very similar in direct $\Upsilon$ decays and in $q\bar{q}$ jets at $\Upsilon$ energy, even though the original processes appeared to be very different in terms of partons (2 quarks at 9.4 GeV, 3 gluons at 9.46 GeV). The interaction through gluons in both cases influences strongly also the non-perturbative hadronization, giving an almost fixed probability of $K^0_s$ per charged particle.

| | $K^0_s$ per event | | | $K^0_s$ per charged particle |
|---|---|---|---|---|
| | PLUTO [130] | ARGUS [118,119] | CLEO [125] | PLUTO [130] |
| ϒ$_{direct}$ data | 0.97 ± 0.22 | 1,033 ± 0.036 | 1.02 ± 0.07 | 0.12 ± 0,03 |
| continuum | 0.73 ± 0.16 | 0.92 ± 0.05 | 0.89 ± 0.04 | 0.11 ± 0,02 |

**Table 5.** Experimental $K^0_s$ production (corrected) in events with 3-gluon or 2-quark jets approximately at the same c.m.s. (continuum data: PLUTO at 9.4 GeV, ARGUS at 9.98 GeV and CLEO at 10.49 GeV).

PLUTO had for the first time in history studied at DORIS an "identified" gluon jet and confirmed (Fig. 12 and Tab. 5) QCD expectations (inbedded also in the simulation of the hadronization) even for the non-perturbative aspects. Many years passed before LEP experiments studied the same physics (having first to identify the gluon jet).

# 6 Confirmations

Every result must be reproduced and confirmed by other experiments. This request for the present subject was fulfilled first at DORIS, later at DORIS II and CESR and (in parallel) at PETRA at high energies.

## 6.1 At DORIS and DORIS II and at CESR

In Tab. 1 we have shown a comparison of the main properties of the ϒ resonance found at DORIS in parallel by the PLUTO, DASP (later DASP2) [20,92,93], DESY-Hamburg-Heidelberg-

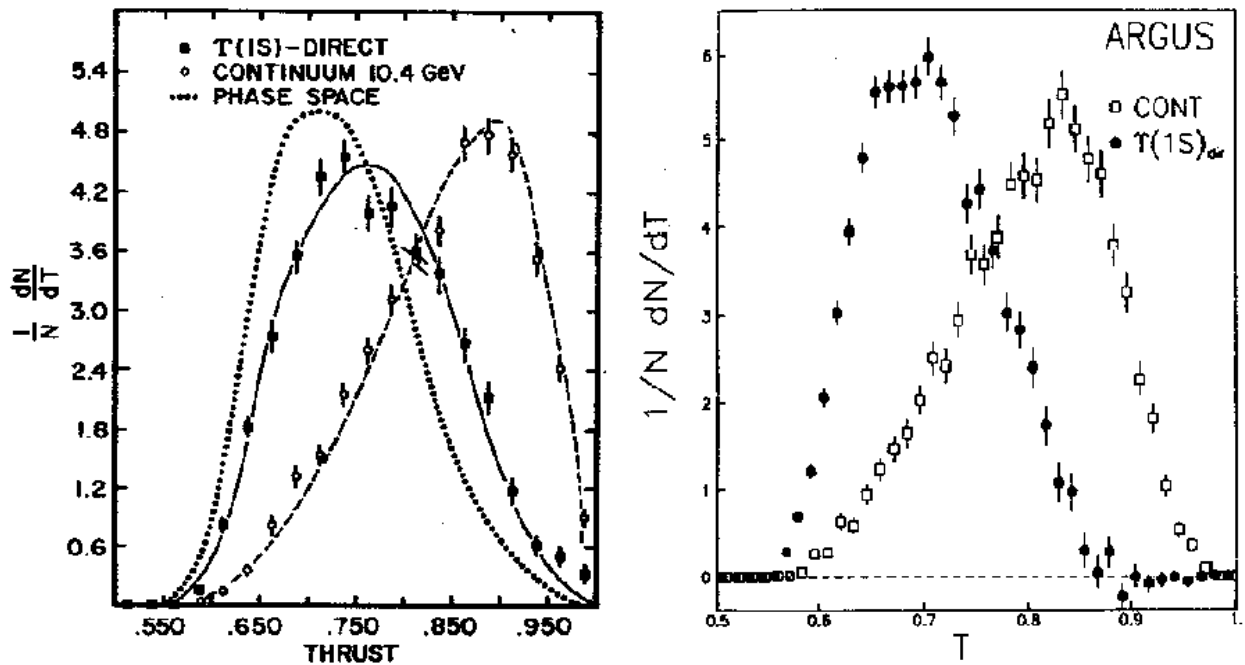

**Fig. 13.** Thrust distributions from CLEO (left, [124]) and from ARGUS (right, [117]) for ϒ(1S) compared to nearby continuum data and (for CLEO, left) to phase space (dotted line), 3-gluon MC (full line) and 2-jet MC (dashed line, at 10.49 GeV).

München (DHHM) [22] experiments. The detectors were different, but the results were comparable. The more sophisticated aspects of the $\Upsilon \rightarrow$3-gluon decays were found only by PLUTO (but <sphericity> and <thrust> were measured also by DHHM Collaboration [22]). Later, using the upgraded DORIS II with better luminosity, two more detectors, LENA [114] and ARGUS with particle identification [117-119][14], explored again the $\Upsilon$ topological variables and studied gluon jets, confirming qualitatively (using only uncorrected thrust data) the PLUTO results. At the same time in the U.S.A. at Ithaca (Cornell Univ.) the sophisticated experiment CLEO [124] and CUSB [120] at CESR also succeeded in confirming the 3-gluon hypothesis. Fig. 13 (right) shows the inclusive distributions of thrust by ARGUS and (left) CLEO, compared with the continuum data and with the three models (3-gluon, 2-jet and phase space).

More in detail, CLEO (Fig. 14) confirmed at $\Upsilon$ the validity of the PLUTO 3g-MC (in principle a rough approximation: gluons hadronizing like quarks, with independent fragmentation): with particle identification, the ratio of multiplicities of charged pions and kaons (black points in Fig. 14) compare well with that kind of MC. (The case of baryons, also shown, is more peculiar of 3 gluon properties, but they contribute only 0.6 particles per event on $\Upsilon_{direct}$.) The PLUTO approximation was then confirmed a posteriori.

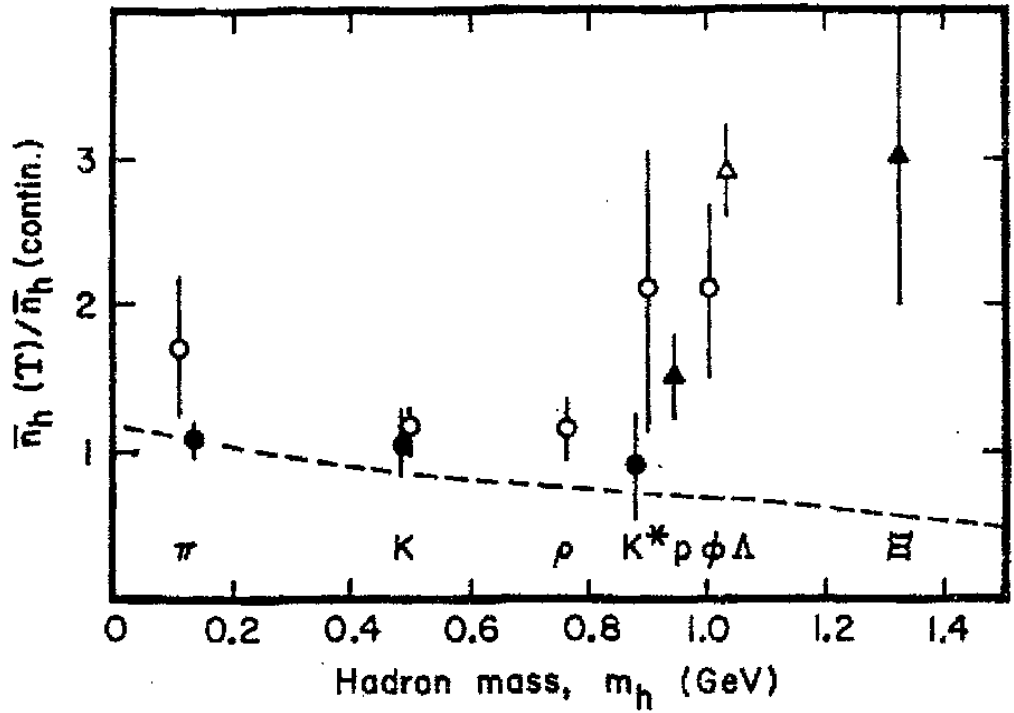

**Fig. 14** Hadronic multiplicity ratios ($\Upsilon_{direct}$/continuum) from CLEO [125]  with particle identification; dashed line: a MC program similar to the PLUTO one (gluons hadronizing like quarks; independent fragmentation).

## 6.2 PLUTO and other experiments at PETRA

PLUTO moved to PETRA in the late summer 1978 and was upgraded with forward spectrometers and muon chambers [94,95]. As being mostly a well tested and known detector, PLUTO was able to publish the first results at the higher c.m.s. energies of 13 and 17 GeV already in February 1979 [94,95]. The longitudinal (<$p_{||}$>) and transverse (<$p_T$>) momenta of jets with respect to the thrust axis were studied and found to increase with $E_{cm}$ in different ways, linearly for <$p_{||}$> and logarithmically or limited for <$p_T$>; which implies a decreasing opening angle for jets. At the Geneva Conference (June 1979) V. Blobel (PLUTO) presented also the new results at 27.4 GeV [47]. The R variable (total hadronic cross section divided by the $\mu^+\mu^-$ cross section) showed an increase compatible (within low statistics) to the opening of the production of a new heavy quark-antiquark pair, of charge |⅓|, as expected from the width of Υ.

[14] PLUTO had moved to PETRA (also at DESY).

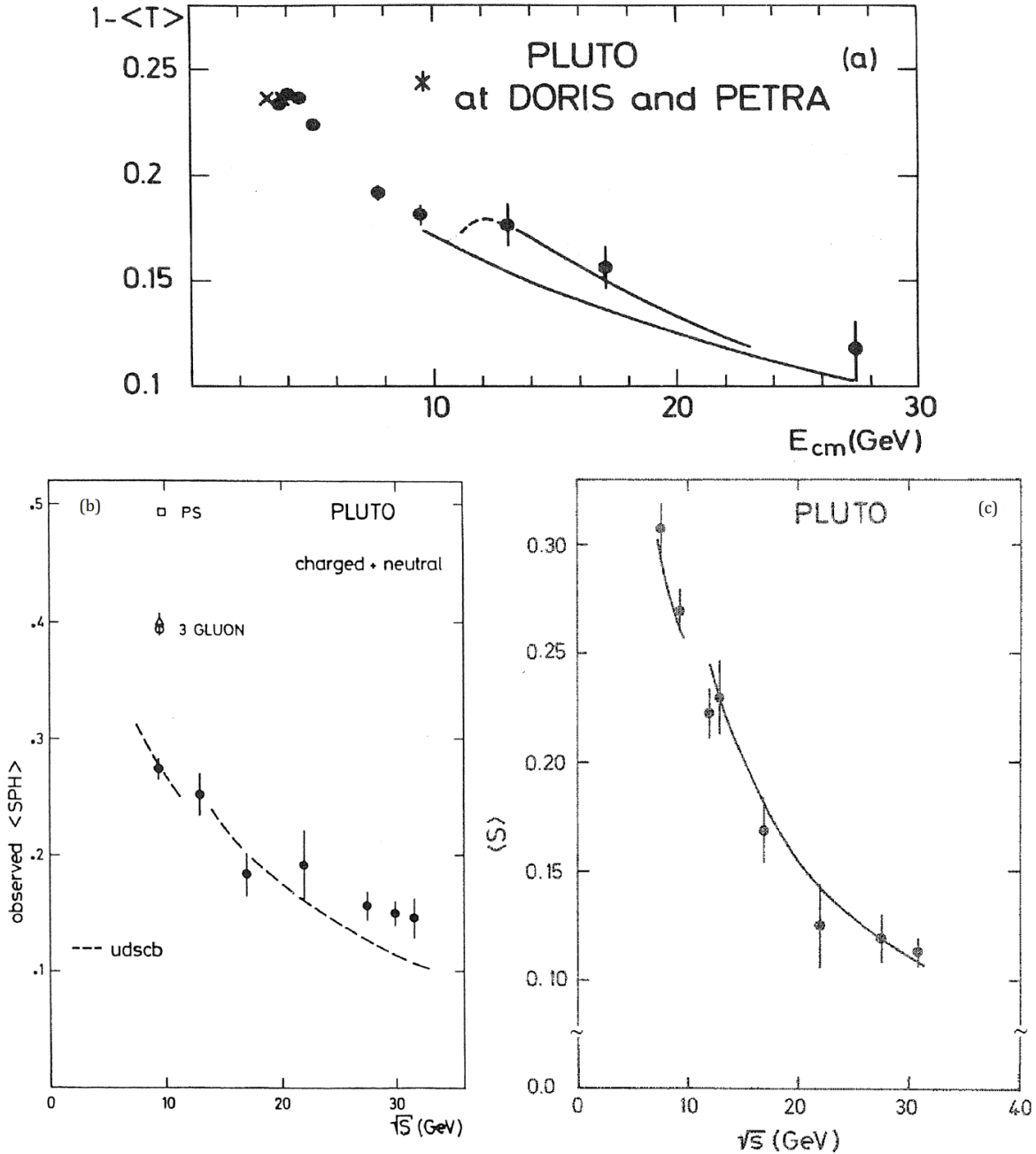

**Fig. 15a.** Evolution with $E_{cm}$ of the inclusive quantity 1-<T>. The solid lines display the expectations with (upper) and without (lower) a new quark with charge ⅓ [94,95,50,132]. The asterisk shows the Υ$_{direct}$ data. **Fig. 15b.** and **c.** The mean sphericity, observed (b) and final (c) [135], using charged and neutral particles, as function of $E_{cm}$ = √s; **b)** compared to the expectations of the 2-jet Field-Feynman MC (dashed line) without gluon bremsstrahlung [53] and **c)** (also charged+neutral) to Hoyer et al. MC (with gluon bremsstrahlung, Fig. 15a). The open points in b) show the Υ data (triangle) compared to the expectations of phase-space (square) and 3-gluon MC (circle).

A more precise comparison was possible with the topological variable 1-<T> as function of $E_{cm}$ [47], according to the expectation of QCD by Ali et al. [132], if $b\bar{b}$ production and decay was included. The data (Fig. 15a) showed an increase with respect to the u, d, s, c quarks only. The opening of the new threshold implied by the very narrow $\Upsilon$ resonance was confirmed by the small step in Fig. 15a (exhibited by PLUTO already in February 1979 [94,95]), fully compatible with the -⅓ charge of the new bottom quark (a ⅔ charge would have produced a much larger step).

At the higher energies of PETRA the evolution with energy of the topological variables was confirmed with higher statistics by PLUTO, using charged and neutral particles (Fig. 15b: note the smaller error bars compared with Fig. 3). The phase space model is completely ruled out and the trend of the data agrees with the Field-Feynman MC with 5 quark flavours (gluon radiation was not yet included). In the inclusive topological variables, at the highest energies (last three points) PLUTO observed an excess with respect to the parton model including fragmentation [53,132]. The evidence for exclusive hard gluon emission (an additional separated jet) was still to be found. The existence of gluons observed in the 3-gluon decays of $\Upsilon$ implied (as in any field theory) the radiation of soft and hard gluons by quarks and then at first a broadening of the jets with increasing energy with respect to the case of no radiation. For instance, $<p_T^2>$ was expected to rise as: $<p_T^2> \sim \alpha_s(s)$ s (where s=$E^2_{cm}$ and $\alpha_s$ is the strong coupling constant, decreasing with log $E_{cm}$), due to the emission of the soft quanta of the strong force (similar to bremsstrahlung in QED) [100]. This was first confirmed inclusively by TASSO at Geneva with charged particles [45.46] and later by PLUTO at Batavia with charged and neutral particles [49]. Figure 16 shows the latter, $<p_T^2>$ dependence on $E_{cm}$ from PLUTO, the data follow the expectation for $q\bar{q} + q\bar{q}g$, (solid line), and rule out the expectation for $q\bar{q}$ only (dashed line).

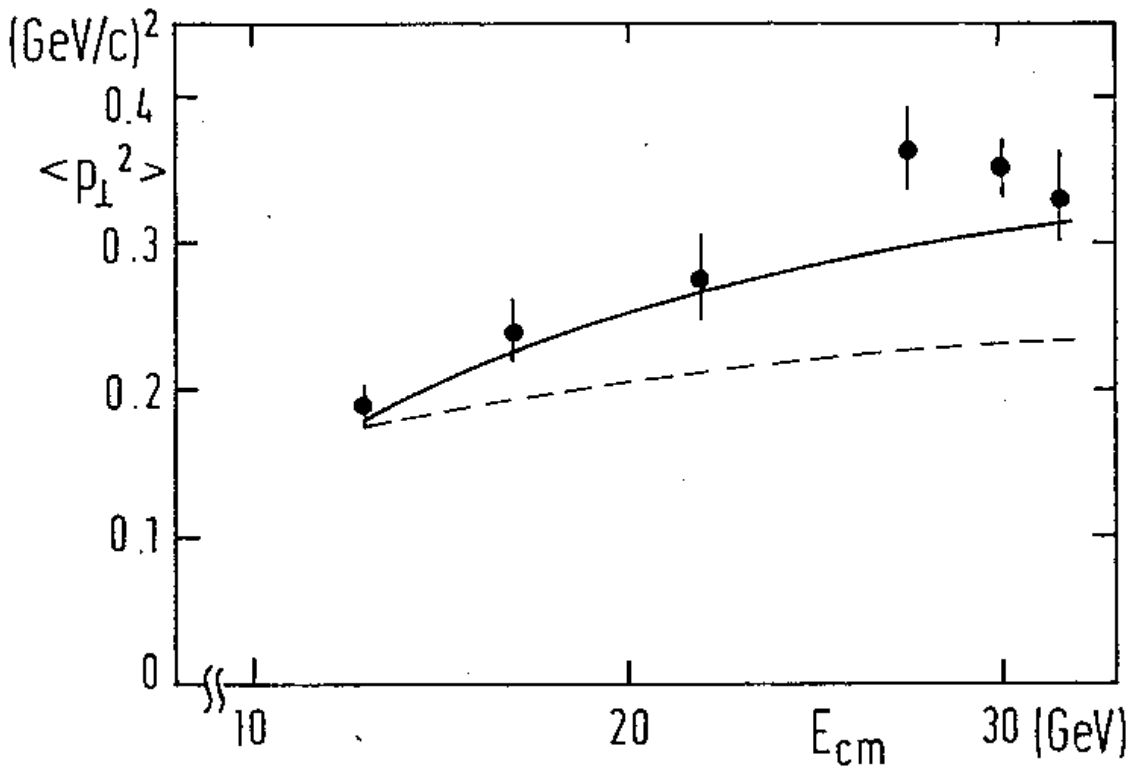

**Fig. 16.** Evolution of $<p_t^2>$ of the charged particles belonging to the fastest jet per event (PLUTO, charged + neutrals thrust axis) with $E_{cm}$, exhibiting jet broadening [49,53]. The dashed line is the MC expectation for $q\bar{q}$ jets (all "flavours", b included) and the solid line is the same plus gluon radiation. [132,102]

In June 1979, B. Wiik (TASSO) exhibited the first evidence of three jet-like events (a single event using only charged particles) at PETRA (Bergen Conference, [51]). Later at Geneva [45] P. Söding (TASSO) showed a few more events; all events were reconstructed yet without energy and momentum conservation.

PLUTO [49,50,53,57], and the other PETRA experiments confirmed [52,54,55] the presence of three exclusive jets. Figure 17 shows a PLUTO 3-jet event; in this case the availability of neutral energy data in the detector gave a significant contribution to reducing the systematic errors (see Fig. 17 from [50]). It should be noted that the gluon bremsstrahlung effect, even at the highest energies of PETRA has only a 10% probability, to be compared with the almost 97% $ϒ_{direct}$ to 3-gluon decay (according to QCD and *confirmed experimentally* by PLUTO). At this point at PETRA, a cross section was not yet measured.

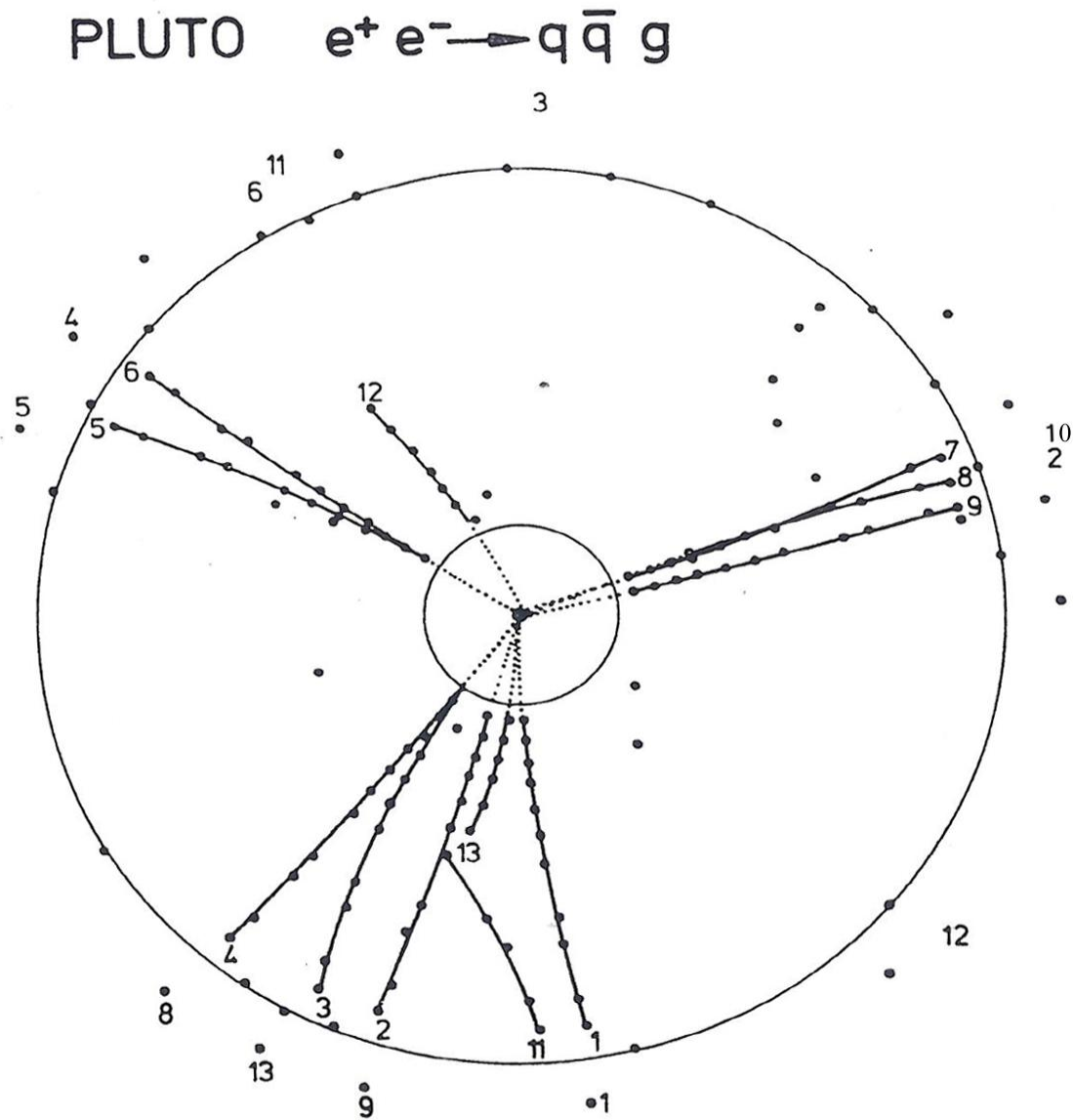

**Fig. 17.** An evident 3-jet event (interpreted as q$\bar{q}$g) from PLUTO data at PETRA: 13 charged tracks (8 vertex fitted) and 5 neutral clusters (out of 13 showers) were reconstructed. (The numbers are just labels: energy of showers and momentum of tracks are not shown here [50,53]).

The gluon jet was not identifiable in $q\bar{q}g$ events at PETRA (even at LEP no experiment has identified event by event the gluon jets). Could gluon spin be determined experimentally as had been done by PLUTO for $\Upsilon \to 3$ gluon decays? PLUTO [57] used the measurement of the differential cross section of the fractional energy of the fastest jet compared with the MC expectation for the fastest parton (Fig. 18). PLUTO measured $\langle x_1 \rangle = 0.893 \pm 0.005$, to be compared with 0.891 for the vector case and 0.871 for the scalar case, and again spin zero was excluded by 4.4 standard deviations. TASSO demonstrated it as well with a different method (angular correlation between the three jet axes) [56].

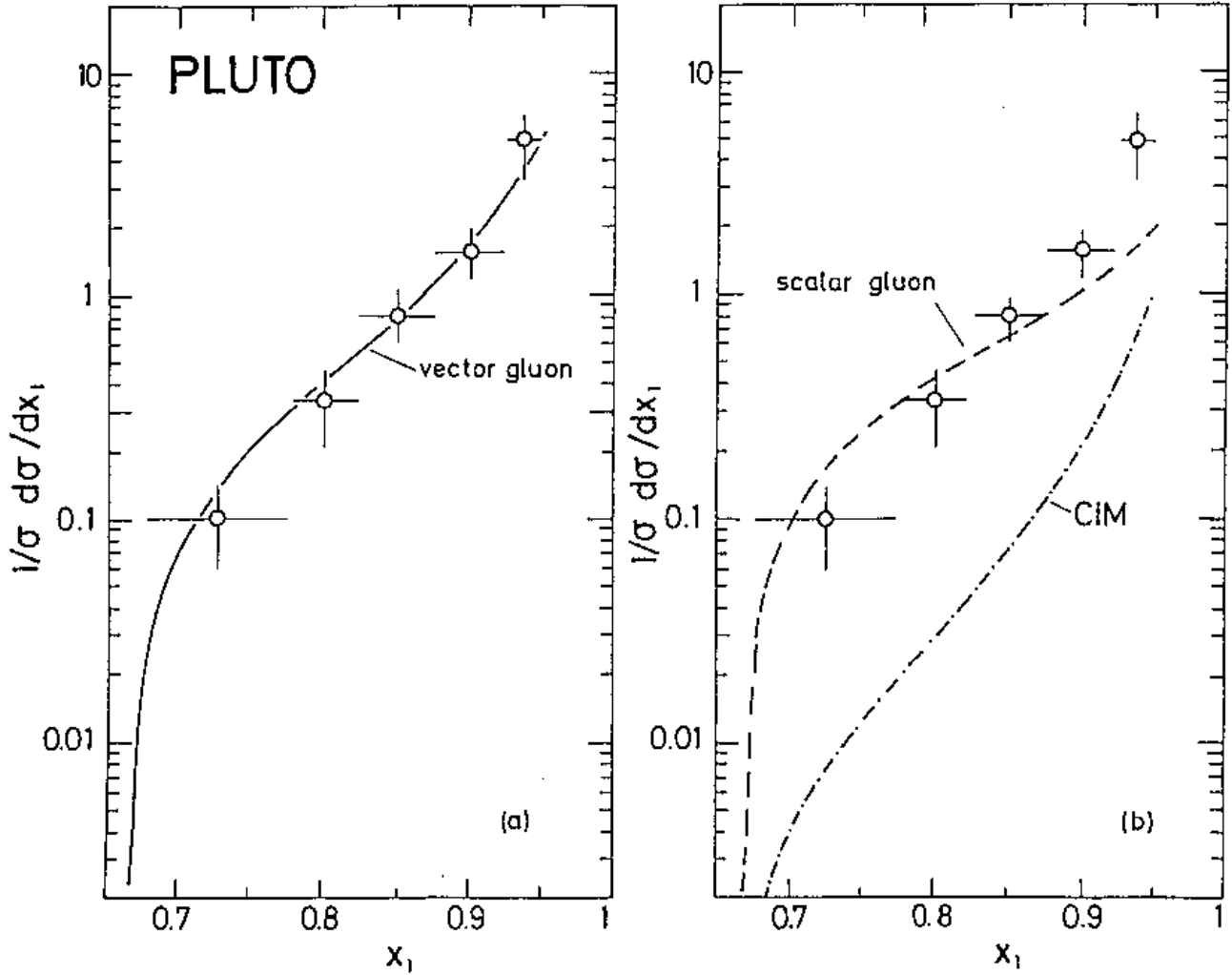

**Fig. 18.** Distribution of the relative energy of the fastest jet ($x_1$) [57]. The data points are corrected for detector acceptance, radiation and hadronization. Solid curve: first order QCD [7,8], dashed curve: scalar gluon hypothesis [129], and dashed-dotted: CIM (Continuous Interchange Model, [134]). The vector gluon hypothesis matches the data also at PETRA (gluon radiation).

Ch. Berger concluded his talk in 1979 [49]: *The evidence for gluons which has been accumulated during the past two years, especially by the work of the PLUTO group on the $\Upsilon$ resonance, gets very strong support from the present experiment.* The gluon discovery by PLUTO in the $\Upsilon$ decays had been confirmed by finding gluon bremsstrahlung at higher energies and measuring again the gluon spin. A striking quantitative comparison by PLUTO with perturbative QCD finally provided the evidence for both the agreement with soft and hard gluon emission and the need for soft gluon resummation [136].

# 7  Summary and Conclusions

The Υ(9.46) is a very narrow heavy resonance ($\Gamma_{ee}$=1.3 KeV [34], $\Gamma_{tot}$=54 KeV), therefore very likely to be a bound state of a new heavy quark-antiquark pair $b\bar{b}$ as confirmed by measuring the excitation curve and the partial and total widths (PLUTO at DORIS and other experiments, see Tab. 1). The opening of the threshold for bottom quark production was confirmed by PLUTO at DORIS by measuring the relative increase of <S> and <1-T> just above the Υ excitation region, showing the development of a step in R above the Υ energy (the increased value being confirmed also at PETRA).

The dominant topology Υ → 3 jets of the hadronic decays of the $b\bar{b}$ ground state Υ(9.46) as expected by QCD, was discovered by PLUTO at DORIS with the data collected in the spring of 1978. The favourable results (initially, using only reconstructed charged tracks giving larger systematic errors for the topological variables; later using also neutrals) were interpreted as 3 gluons, because of the 3-jet topology, of the slightly larger multiplicity and of the convincing agreement with the 3-gluon dynamics. The 3-jet topology was shown first independently by event <$p_T$>, <$p_{out}$>, momentum distribution (see Chap. 5.5 and Tab. 4) and by average values of topological variables (see Chap. 5.2 and Tab. 3), especially first by sphericity (see Fig 3), and later by detailed comparison of differential distributions with models (see Chap. 5.4); it could not come from 3 quarks because Υ is a neutral boson, like the virtual photon from $e^+e^-$ annihilation, nor from the $q\bar{q}g$ having too little energy left for hard bremsstrahlung as a gluon jet in such a large fraction of events. Moreover, the Υ→ γgg decay fraction was expected to be only about 3%.

The QCD 3-gluon matrix element describing the Υ decay features (Fig. 7) favours the appearance of one energetic jet in the final state (of <$E_1$> = 4.2 GeV, as a quark jet at 8.4 GeV c.m.s. energy), determining in every event the thrust or triplicity axis direction (Fig. 8). It manifested itself in PLUTO also as an event <$p_T$> of 0.34±0.01 GeV, the same as that of 2 quark jets in the continuum of 0.33±0.01 GeV, both uncorrected but measured in the same way with large acceptance (Fig. 11 and Tab. 4). This "jettiness" (helped by the well separated second energetic jet of <$E_2$> = 3.4 GeV) was the peculiar feature found experimentally without use of models that allowed PLUTO to give a physical meaning to the thrust axis and to see the correlation with the beam axis (Fig. 10) and to find out that gluons at that energy fragment very similar to quarks and therefore, of course, in a "jetlike" fashion (Tab. 4). Without those jets (modelled with independent quark fragmentation found experimentally) the PLUTO data would have widely disagreed with any model.

In detail the 3-jet topology, indicated by only comparing observed data at Υ and nearby continuum in a model independent way by looking at event characteristics like jettiness (<$p_T$>), more than 2-jets (momentum slope and <S>) and flatness (<$p_{out}$>), was confirmed by comparing the differential distribution of topological variables constructed using all the detected particles and averaged over the plenty of events, with the expectations of different models. More than 15 different variables (including dynamic ones) confirmed the decay topology and the spin-parity $1^-$ of the gluons already in the first half of 1979. As said by Koller and Walsh [9-11]: "*If the 3-gluon jet decay of a heavy $Q\bar{Q}$ state is found, it will in our opinion provide a striking confirmation of QCD*". As J. Ellis anticipated [137]: "*topological variables as antenna patterns could be … used to extract statistical evidence for gluon radiation, even if individual 3-jet events could not be distinguished*". This is what PLUTO did using the Υ → 3-gluon decay (with gluons having a hard energy spectrum, contrary to that of gluon  bremsstrahlung).

Since 15 topology variables in the data (although many of them correlated) agreed with that hadronization of 3 jets, interpreted as gluons, this is also *the first confirmation of gluon jets*.

In June 1979 [43] the 3-jet topology was studied in detail experimentally; the matrix element density distributions were measured and agreed quantitatively with the 3-gluon hypothesis and the spin $1^-$ of the original parton of the jets was demonstrated. All other possible model explanations were rejected and an important confirmation of QCD was provided; *the forming and hadronization of "identified" gluon jets were observed and studied for the first time.*

This splendid result was soon confirmed by PLUTO itself, transferred to PETRA in the fall of 1978, at a factor two (and then up to four) higher c.m.s. energies. At PETRA, PLUTO was the only experiment which had analyzed data from 3 GeV (at DORIS) to 32 GeV c.m.s. energy with the same detector, at the J/ψ and Υ and in the extensive e$^+$e$^-$ continuum.  The advantage of being ready (with a well known detector) to operate from the first day allowed PLUTO to find the first hadronic events (November 1978) and to achieve a study of the jets at PETRA in February 1979 [35] and contributed to minimizing the systematic uncertainties. As TASSO did in June 1979 [45,46], PLUTO [49] provided further evidence for gluon emission inclusively and quantitatively in all hadronic events, by confirming the jet broadening due to gluon radiation as expected by QCD (plus hadronization). The exclusive 3-jet topology was then found in a fraction of events by scanning (Fig. 17); the much less frequent (compared to the Υ→3-gluon decays) visual topology found in a few events being interpreted as q$\bar{q}$ + hard gluon radiation. PLUTO used from the beginning both charged and neutral particles, as did MARK-J (but without a magnetic field) and a month later JADE. The TASSO results used only charged particles.  The earlier results of PLUTO at DORIS were also a stimulus to the new experiments to search for the 3-jet topology in the larger phase space and easier kinematics of the higher energy PETRA machine. Jets (in analogy to molecules[15]) were directly visible here in a fraction of selected events. PLUTO had observed the effect of gluon jets one year earlier, statistically and quantitatively in Υ$_{direct}$ hadronic decays.

We have critically recalled and summarised the main results of the PLUTO communications and publications in the years 1978 and first half of 1979 (and later, for completeness) on the Υ(9.46) to 3-jet final state interpreted as 3 gluon jets. We have demonstrated that the 3-gluon MC was able to describe all possible inclusive variables as well as the proposed parton dynamics even after hadronization (Fig. 8 mostly and Figs. 4,6,9,10,12) and explored then for the first time the hadronization of the "identified" gluons exhibiting jet features (Fig. 11 and Tab. 4), and all together, as expected by QCD, a larger multiplicity. Agreement with the expected QCD matrix element and the gluon $1^-$ spin-parity was demonstrated as well. We have shown that no other reasonable model was (and is) able to describe the data (Figs. 4,5,6,10).

Although in lepton-nucleon scattering there had been indications (1970-1976) of the existence of the gluon, such sufficient and necessary conditions like reconstructed in Υ→3-jet→3-gluon decays by Pluto had not been provided.

The confirmation of the PLUTO results on the Υ was achieved at DORIS I, DORIS II and CESR by other experiments, some with significantly better detectors. Also the PLUTO MC model for gluon hadronization was confirmed a posteriori (Fig. 14). At PETRA, the observation of jet broadening, the evidence of gluon bremsstrahlung found by 4 parallel experiments (including PLUTO) and the measurement of the gluon spin confirmed the existence and properties of QCD gluons that PLUTO had already put in evidence at lower energy in Υ decays.

---

[15] As Perrin confirmed the existence of invisible molecules by the brownian motion of visible grains hit by a large number of invisible molecules and by the Einstein theory of it (an analogy suggested by S. Brandt (PLUTO, TASSO) in episode 94 in his recent historical book [6]).

**Acknowledgements**. We thank the PLUTO members Hinrich Meyer for his support and encouragement and for emphasizing the importance of detection of the neutral particles, Gerhard Knies for recalling the quasi-two-jets event topology, Claus Grupen and Robin Devenish for the carefully reading of the first draft (DESY report 10-130). We thank Ahmed Ali and Mario Greco for reading the first draft and for discussions on various aspects of the ϒ decay, QCD and jet models. The PLUTO Collaboration is warmly acknowledged for the results.

**Note added in proof**

In their new historical review on "*JETS and QCD: a historical review of the discovery of the quark and gluon jets and its impact on QCD*" [140], Ali and Kramer deal also with our subject, ϒ(9.46) and the gluon discovery, within the Chapters 1 "Introduction" (see page 247), 4 "Gluon jets in ϒ decays" (see page 264) and 8 "Summary" (see page 311). Their evaluation is summarized at page 311: "*A clear three-jet topology using en vogue jet definitions was not established in ϒ(9.46) decays for lack of* energy *[24].*" Actually in the present paper we advocate the three-jet topology both by comparing many differential distributions of topological variables with the three-gluon MC expectations and by comparing average event variables on resonance with the same ones in the continuum. ϒ(1s) was expected by QCD to decay into 3 gluons hadronizing in jets of average energy $<E_1>$=4.2, $<E_2>$=3.4 and $<E_3>$=1.8 GeV, compared to PLUTO experimental values 4.08±0.01, 3.38±0.02 and 2.00±0.02 GeV, respectively [97]. It must be recalled that jets were observed at SPEAR at 6.2 and 7.4 GeV c.m.s. energies [65] (later even extended down to 4.8 GeV c.m.s. energy [66,67]), corresponding to half of it for single jets (2.4, 3.1 and 3.7 GeV): only $<E_3>$ is smaller than those, but it is fixed just by energy and momentum conservation. SPEAR studied sphericity (as PLUTO did later) and was credited for having discovered quark jets. There was then no lack of energy at ϒ(9.46), exhibiting jets in the same range of energy as SPEAR or larger, and indeed Ali and Kramer recognize that the jets were measurable at page 264 of their text : "*Average energies of the three partons were measured as  $<E_1>$~4.1 GeV for the most energetic of the three gluons, with the other two having energies $<E_2>$~3.4 and $<E_3>$~2.0 GeV, respectively [111], in approximate accord with the lowest order QCD matrix elements.*" Also the three angles between the jets were measured to be in accord (Fig. 9). A parton can be observed and measured only as a jet of hadrons, so that the previous quoted sentence is referred necessarily to the 3 gluon jets. They continue: "*However, only the fastest of the three partons yielded a collimated jet of hadrons and its detailed phenomenological profile was studied by PLUTO [111,112].*" Ali and Kramer admit here implicitly that PLUTO has identified for the first time single gluon jets. [85] corresponds to the present paper; check our Chapter 5.4: "*The exclusive three gluon dynamics.*". If PLUTO had demonstrated the ϒ main decay to be in approximate accord with QCD and identified *at least* a single hadron jet per event in a wide range of energies (Fig. 9, $x_1$), with the event $<p_T>$ as small as for quark jet events measured by the same experiment, and if this fastest jet was found by PLUTO to have the gluon spin and the events had a larger charged multiplicity than the corresponding experimental quark jets, everyone can ask himself what could be the remaining objects, measured to have energies (and angles) "*in approximate accord with the lowest order QCD matrix elements*" for 3 gluons, in case PLUTO had indeed not observed the three-jet topology and found the three jet axes?  The only remaining conclusion is that we have observed three gluon jets.

As Gunter Wolf (TASSO,DESY) stated at Erice 1978 [27]: "*All aspects investigated so far are in accord with the assumption that the  ϒ decays via a three gluon intermediate state. This is in strong support of QCD.*"

## Appendix: PLUTO Collaboration

At the time of [34] the authors were:

Ch. Berger, W. Lackas, F. Raupach, and W. Wagner (Aachen);

G. Alexander, J. Bürger, L. Criegee, H.C. Dehne, K. Derikum, R. Devenish, G. Flügge, G. Franke, Ch. Gerke, E. Hackmack, P. Harms, G. Horlitz, Th. Kahl, G. Knies, E. Lehman, B. Neumann, B. Stella, R.L. Thompson, U. Timm, P. Waloschek, G.G. Winter, S. Wolff, and W. Zimmermann (DESY);

O. Achterberg, V. Blobel, L. Boesten, H. Daumann, A.F. Garfinkel, H. Kapitza, B. Koppitz, W. Lührsen, R. Maschuw, H. Spitzer, R. van Staa, and G. Wetjen (Hamburg II Inst.);

A. Bäcker, S. Brandt,  C. Grupen, H.J. Meyer, and G. Zech (Siegen);

K. Daum, H. Meyer, O. Meyer, M. Rössler, and K. Wacker (Wuppertal).

(The number of authors, according to W. Wagner, fulfilled the law $\langle n_{auth} \rangle = 26 + 2.2\ E_{cm}(GeV)$.)